\documentclass[reqno,11pt]{amsart}
\usepackage{hyperref}
\usepackage{amssymb}
\usepackage[mathscr]{eucal}
\usepackage{graphicx, esint}
\numberwithin{equation}{section}
\allowdisplaybreaks[3]

\title[Entanglement and Second Quantization]{ \vspace*{-0.9cm}
Entanglement and Second Quantization in the Framework of the Fermionic Projector}

\author[F.\ Finster]{Felix Finster \\ \\ November 2009
\vspace*{-1em}}
\thanks{Supported in part by the Deutsche Forschungsgemeinschaft.}
\address{Fakult\"at Mathematik \\ Universit\"at Regensburg \\ D-93040 Regensburg \\ Germany}
\email{Felix.Finster@mathematik.uni-regensburg.de}
\newtheorem{Def}{Definition}[section]
\newtheorem{Thm}[Def]{Theorem}
\newtheorem{Prp}[Def]{Proposition}
\newtheorem{Lemma}[Def]{Lemma}
\newtheorem{Remark}[Def]{Remark}

\newtheorem{Example}[Def]{Example}

\newcommand{\Thanks}{\vspace*{.5em} \noindent \thanks}
\newcommand{\beq}{\begin{equation}}
\newcommand{\eeq}{\end{equation}}
\newcommand{\Proof}{\begin{proof}}
\newcommand{\QED}{\end{proof} \noindent}
\newcommand{\QEDrem}{\ \hfill $\Diamond$}
\newcommand{\spc}{\;\;\;\;\;\;\;\;\;\;}
\newcommand{\la}{\langle}
\newcommand{\ra}{\rangle}
\DeclareMathOperator{\bra}{\,<\!}
\DeclareMathOperator{\ket}{\!>\,}
\newcommand{\C}{\mathbb{C}}
\newcommand{\R}{\mathbb{R}}
\newcommand{\1}{\mbox{\rm 1 \hspace{-1.05 em} 1}}
\newcommand{\Z}{\mathbb{Z}}
\newcommand{\N}{\mathbb{N}}
\renewcommand{\H}{\mathscr{H}}
\newcommand{\F}{\mathscr{F}}
\newcommand{\FHF}{\mathscr{F}^\text{\tiny{HF}}}

\newcommand{\cO}{\mathcal{O}}

\DeclareMathOperator{\Tr}{Tr}

\DeclareMathOperator{\sign}{sign}
\DeclareMathOperator{\diag}{diag}

\newcommand{\U}{\text{\rm{U}}}
\newcommand{\psiu}{\psi^\uparrow}
\newcommand{\psid}{\psi^\downarrow}
\renewcommand{\O}{{\mathscr{O}}}
\newcommand{\Sl}{\mbox{$\prec \!\!$ \nolinebreak}}
\newcommand{\Sr}{\mbox{\nolinebreak $\succ$}}
\newcommand{\SU}{{\rm{SU}}}
\newcommand{\cl}{\text{class}}
\begin{document}
\maketitle

\begin{abstract}
A method is developed for realizing entangled states and
general second quantized fermionic and bosonic fields in the framework of the fermionic projector.
\end{abstract}

\tableofcontents

\section{Introduction}
In~\cite{PFP, srev} it was proposed to formulate physics based on a new action principle in space-time.
One difference of this approach to standard quantum field theory is that a many-particle state no
longer corresponds to a vector in the fermionic Fock space,
but instead it is described by the so-called fermionic projector, an operator which acts on the
one-particle Hilbert space (or more generally on an indefinite inner product space spanned by
the one-particle wave functions). Another difference is that the bosonic fields obtained in the
so-called continuum limit are only classical.
Due to these differences, it is not at all obvious whether the fermionic projector
can account for all quantum effects observed in nature. More specifically, is it
possible to describe {\em{entanglement}}? Can one reproduce the
effects of {\em{second quantized fields}}?

In this paper, we shall analyze these questions in detail. We
will show that it is indeed possible to describe entangled states as well as general second quantized
bosonic and fermionic fields in the framework of the fermionic projector.
This is achieved by introducing the physical concept of a
{\em{microscopic mixing of decoherent subsystems}}.
The physical picture is that space-time is not smooth on the microscopic scale
(typically thought of as the Planck scale), but has a non-trivial microstructure.
Homogenizing this microstructure, we obtain an effective description of the system
by a vector in the Fock space. To make this picture precise, we use the fact that in the framework of the
fermionic projector, the usual topological and causal structure of Minkowski space is
not a-priori given, but it is induced on the space-time points by the states of the
fermionic projector.
Thus by bringing the wave functions between certain pairs of space-time points
``out of phase'', we obtain decoherence effects which result in a decomposition of
the whole system into subsystems between which the usual causal relations are no longer valid.
This makes it possible to realize many independent physical systems simultaneously in
one spacetime, in such a way that homogenizing on the microscopic scale leads to
an effective ``superposition'' of the subsystems. For technical simplicity, we will
describe the microscopic mixing by localizing the subsystems in disjoint
space-time regions (see Figure~\ref{fig2} on page~\pageref{fig2}). But one can also
think of the subsystems as being delocalized, similar as if one combines several images in a
single hologram (see Section~\ref{secqurem}).

In order to make the paper self-contained easily accessible, we begin in Chapter~\ref{secPFP}
with a brief outline of the fermionic projector approach. In Chapter~\ref{sec2},
we recall the basics on entanglement and Fock spaces and work out the connection to projectors
in the one-particle Hilbert space.
In Chapter~\ref{secmix} we describe entanglement using the concept of microscopic mixing
and by introducing a decoherence between space-time regions.
As is worked out in Chapter~\ref{secbosonic}, this notion of decoherence also makes it possible to
describe second quantized bosonic fields.

\section{An Outline of the Fermionic Projector Approach} \label{secPFP}
In this chapter we give a brief introduction to the framework of the fermionic projector,
outlining a few ideas, methods and results. For details we refer to~\cite{PFP, discrete, sector}
or to the review papers~\cite{lrev, srev}.

\subsection{An Action Principle for Fermion Systems in Discrete Space-Time} \label{secdiscrete}
In the fermionic projector approach, the physical equations are formulated intrinsically
in a discrete space-time. To introduce the basic framework, we let~$(H, \bra .|. \ket)$
be a finite-dimensional complex vector space endowed with an indefinite inner product
(thus~$\bra .|. \ket$ is non-degenerate, but not positive definite).
Next, we let~$M = \{1, \ldots, m\}$ be a finite set. To every point~$x \in M$ we associate a
projector $E_x$ (a projector in~$H$ is defined just as in Hilbert spaces as a linear
operator which is idempotent and self-adjoint). We assume that these projectors are orthogonal and
complete in the sense that
\beq \label{oc}
E_x\:E_y \;=\; \delta_{xy}\:E_x \spc {\mbox{and}} \spc
\sum_{x \in M} E_x \;=\; \1\:.
\eeq
Furthermore, we assume that the images~$E_x(H) \subset H$ of these
projectors are non-de\-ge\-ne\-rate subspaces of~$H$, which
all have the same signature~$(2,2)$.
The points~$x \in M$ are called {\em{discrete space-time points}}, and the corresponding
projectors~$E_x$ are the {\em{space-time projectors}}. The
structure $(H, \bra .|. \ket, (E_x)_{x \in M})$ is called {\em{discrete space-time}}.
The particles of our system are described by one
more projector~$P$ in~$H$, the so-called {\em{fermionic projector}}, which
has the property that its image~$P(H)$ is a
negative definite subspace of~$H$.
The resulting system~$(H, \bra .|. \ket, (E_x)_{x \in M}, P)$ is referred to as
a {\em{fermion system in discrete space-time}}.

Let us briefly discuss these definitions and introduce a convenient notation.
The vectors in the image of~$P$ have the interpretation as the
occupied fermionic states of our system, and thus the rank of~$P$
gives the {\em{number of particles}} $f := \dim P(H)$.
The space-time projectors~$E_x$ can be used to project vectors of~$H$
to the subspace~$E_x(H) \subset H$. Using a more graphic notion, we
also refer to this projection as the {\em{localization}} at the space-time point~$x$.
For the localization of a vector~$\psi \in \H$ we use the short notation
\beq \label{wave}
\psi(x) := E_x\,\psi
\eeq
and refer to~$\psi(x)$ as the corresponding {\em{wave function}}.
Having the connection to relativistic quantum mechanics in mind (see Section~\ref{seccontinuum}
below), we refer to~$E_x(H)$ as the {\em{spinor space}}
at~$x$ and denote it by~$S_x$. It is endowed with the inner product~$\bra .| E_x . \ket$
of signature~$(2,2)$, which we also denote by~$\Sl .|. \Sr$. Using the relations~\eqref{oc},
we can then write
\beq \label{stipdiscrete}
\bra \psi | \phi \ket = \sum_{x \in M} \Sl \psi(x) | \phi(x) \Sr \:.
\eeq
The localization of the fermionic projector is denoted by~$P(x,y) := E_x\,P\,E_y$.
This operator maps~$S_y\subset H$ to~$S_x$, and we usually
regard it as a mapping between these subspaces,
\[ P(x,y) = E_x\,P\,E_y \::\: S_y \rightarrow S_x \:. \]
Again using~\eqref{oc}, we can write the wave function corresponding to~$P \psi$ as follows,
\[ (P\psi)(x) \;=\; E_x\: P \psi \;=\; \sum_{y \in M} E_x\,P\,E_y\:\psi
\;=\; \sum_{y \in M} (E_x\,P\,E_y)\:(E_y\,\psi) \:, \]
and thus
\beq \label{diskernel}
(P\psi)(x) \;=\; \sum_{y \in M} P(x,y)\: \psi(y)\:.
\eeq
This relation resembles the representation of an operator with an integral kernel, and therefore we call~$P(x,y)$ the {\em{discrete kernel}} of the fermionic projector.
Finally, it is often useful to choose an orthonormal basis~$\psi_1, \ldots, \psi_f$ of~$P(H)$
(i.e.\ $\bra \psi_i | \psi_j \ket = - \delta_{ij}$).
Then the fermionic projector and its discrete kernel can be written in bra/ket-notation as
\beq \label{Pdiscrete}
P = -\sum_{j=1}^f |\psi_j \ket \!\bra \psi_j | \qquad \text{and} \qquad
P(x,y) = -\sum_{j=1}^f |\psi_j(x) \Sr \Sl \psi_j(y) |\:.
\eeq

In order to introduce our action principle, for any~$x, y \in M$
we define the {\em{closed chain}}~$A_{xy}$ by
\beq \label{cc}
A_{xy} \;=\; P(x,y)\: P(y,x) \::\: S_x \rightarrow S_x \:.
\eeq
We denote the eigenvalues of~$A_{xy}$ counted with algebraic multiplicities
by~$\lambda_1,\ldots,\lambda_{4}$ and define the {\em{spectral weight}}~$|A_{xy}|$ by
\[ |A_{xy}| \;=\; \sum_{j=1}^4 |\lambda_j|\:. \]
Similarly, one can take the spectral weight of powers of~$A_{xy}$.
Our action principle is to
\beq \label{action}
\begin{split}
\text{minimize the action} \quad {\mathcal{S}}[P] &:= \sum_{x,y \in M} |A_{xy}^2| \\
\text{under the constraint} \quad {\mathcal{T}}[P] &:= \sum_{x,y \in M} |A_{xy}|^2
= {\mbox{const}} \:,
\end{split}
\eeq
where we consider variations of the fermionic projector, keeping the number of particles
as well as discrete space-time fixed.
In~\cite{discrete} it is shown that minimizers of this nonlinear variational principle exist.
For a discussion of the underlying physical principles see~\cite[Section~2]{rev}.

\subsection{The Correspondence to Minkowski Space} \label{secminkowski}
At first sight, the above setting seems inappropriate for physical applications because important
structures like the notion of causality, topology and metric of space-time, gauge fields, etc.,
are missing. However, the idea is that these additional structures arise as a consequence of
a self-organization of the states of the fermionic projector as described by our action principle.
More specifically, for a given minimizing fermionic projector~$P$, its discrete kernel~$P(x,y)$
induces relations between the points~$x$ and~$y$, which can be used to introduce additional
structures in space-time. This mechanism is referred to as {\em{spontaneous structure formation}}.
The first rigorous result on spontaneous structure formation was obtained in~\cite{osymm},
where it is shown that the permutation symmetry of the space-time points is spontaneously broken,
giving rise to a non-trivial outer symmetry group. More detailed information is obtained by
analyzing the eigenvalues of the closed chain:

\begin{Def} {\bf{(causal structure)}} \label{defcausal}
Two space-time points~$x,y \in M$ are called {\bf{timelike}} separated if
the spectrum of~$A_{xy}$ is real. The points are
{\bf{spacelike}} separated if the spectrum of~$A_{xy}$ forms two complex
conjugate pairs having the same absolute value. In all other cases, the two
points are {\bf{lightlike}} separated.
\end{Def} \noindent
Furthermore, from the discrete kernel~$P(x,y)$ one can deduce objects of differential geometry
like the spin connection and curvature (see~\cite{lqg}).
Finally, the papers~\cite{reg, vacstab} show that in the limit of an infinite number of particles
and space-time points, our action principle admits minimizers which
are regularizations of vacuum Dirac sea structures in Minkowski space (see also~\cite{lrev}).
In this limit, the wave function~$\psi(x)$, \eqref{wave}, goes over to a
four-component Dirac spinor. The spin scalar product becomes~$\Sl \psi(x) | \phi(x) \Sr
= \overline{\psi(x)} \phi(x)$, where~$\overline{\psi} = \psi^\dagger \gamma^0$ is the usual
adjoint spinor of Dirac theory. Discrete space-time goes over to a space-time continuum~$M=\R^4$,
and the sum in~\eqref{stipdiscrete} becomes a space-time integral,
\begin{equation} \label{stip}
\bra \psi | \phi \ket := \int_M \Sl \psi(x) | \phi(x) \Sr \: d^4x\:.
\end{equation}
In the simplest case of one sea, the discrete kernel corresponds to the Fourier integral
of the lower mass shell,
\beq \label{Psea}
P^\text{sea}(x,y) \;=\; \int \frac{d^4k}{(2 \pi)^4}\: (k_j \gamma^j+m)\:
\delta(k^2-m^2)\: \Theta(-k^0)\: e^{-ik(x-y)} \:,
\eeq
where~$\Theta$ is the Heaviside function (more generally, one can take sums
or direct sums of Dirac seas to describe different types of elementary particles;
see~\cite[\S5.1]{PFP}).
Here the exponent~$e^{-ik(x-y)}$ involves the Minkowski metric. Even more, \eqref{Psea}
determines the Minkowski metric and can be used define it. If this is done, the causal structure
corresponding to the Minkowski metric indeed agrees with the spectral definition
in Def.~\ref{defcausal} (for details see~\cite[Section~3]{rev}).
Moreover, one can introduce all the familiar objects of Dirac theory. For example,
the non-negative quantity~$\Sl \psi | \gamma^0 \psi \Sr$ has the interpretation as the
probability density of the Dirac particle. Polarizing and integrating over space yields the scalar product
\begin{equation} \label{print}
(\psi | \phi) = \int_{t=\text{const}} \Sl \psi(t,\vec{x}) \,|\, \gamma^0 \phi(t,\vec{x}) \Sr \:d \vec{x}\:.
\end{equation}
For solutions of the Dirac equation, current conservation implies that this 
scalar product is time independent.

\subsection{The Continuum Limit, a Formulation of Quantum Field Theory} \label{seccontinuum}
The above correspondence to vacuum Dirac sea structures can also be used to analyze our
action principle for interacting systems in the so-called {\em{continuum limit}}
(for details see~\cite[Chapter~4]{PFP} and~\cite{sector}). We now outline a few ideas and
constructions needed later on.
First, it is helpful to observe that the vacuum fermionic projector~\eqref{Psea}
is a solution of the Dirac equation~$(i \gamma^j \partial_j - m) P^\text{sea}(x,y)=0$.
To introduce the interaction, we replace the free Dirac operator by a more general Dirac operator,
for example involving gauge potentials or a gravitational field. Thus, considering for simplicity
an electromagnetic potential~$A$, we demand that
\beq \label{DiracP}
\left( i \gamma^j (\partial_j - ie A_j) - m \right) P(x,y) = 0 \:.
\eeq
Moreover, we introduce particles and anti-particles by occupying (suitably normalized)
positive-energy states and removing states of the sea,
\beq \label{particles}
P(x,y) = P^\text{sea}(x,y)
-\frac{1}{2 \pi} \sum_{k=1}^{n_f} |\psi_k(x) \Sr \Sl \psi_k(y) |
+\frac{1}{2 \pi} \sum_{l=1}^{n_a} |\phi_l(x) \Sr \Sl \phi_l(y) | \:.
\eeq
Using the so-called causal perturbation expansion and light-cone expansion,
the fermio\-nic projector can be introduced via~\eqref{DiracP} and~\eqref{particles}.

It is important that our setting so far does not involve the field equations; in particular,
the electromagnetic potential in the Dirac equation~\eqref{DiracP} does not need to satisfy
the Maxwell equations. Instead, the field equations should be derived from
our action principle~\eqref{action}. Indeed, analyzing the corresponding Euler-Lagrange equations,
one finds that they are satisfied only if the potentials in the Dirac equation satisfy certain
constraints. Some of these constraints are partial differential equations involving
the potentials as well as the wave functions of the particles and anti-particles in~\eqref{particles}.
In~\cite{sector}, such field equations are analyzed in detail for a system involving an
axial field. In order to keep the setting as simple as possible, we here consider the analogous
field equation for the electromagnetic field
\beq \label{Maxwell}
\partial_{jk} A^k - \Box A_j = e \sum_{k=1}^{n_f} \Sl \psi_k | \gamma_j \psi_k \Sr
-e \sum_{l=1}^{n_a} \Sl \phi_l | \gamma_j \phi_l \Sr \:.
\eeq
With~\eqref{DiracP} and~\eqref{Maxwell}, the interaction as described by the
action principle~\eqref{action} reduces in the continuum limit to the
{\em{coupled Dirac-Maxwell equations}}.
The many-fermion state is again described by the fermionic projector, which is built up
of {\em{one-particle wave functions}}. The electromagnetic field is merely a
{\em{classical bosonic field}}.

For the considerations in Chapters~\ref{secmix} and~\ref{secbosonic}, it is important to keep in
mind that in the framework of the fermionic projector,
space-time is {\em{not smooth}} on the microscopic scale, but it has
an underlying discrete structure. The dynamics is described intrinsically in discrete space-time
by the action principle~\eqref{action}. A minimizing fermionic projector has a rich microscopic
structure from which one can deduce notions which have a correspondence to macroscopic
physics. In the 
continuum limit, the causal and metric
structure of Minkowski space can be recovered from the fermionic projector
using the notion of Definition~\ref{defcausal}. Thus we can say that the
{\em{wave functions}} of the Dirac particles (also taking into account the states of the Dirac sea)
{\em{generate the causal and geometric structure of space-time}}. This observation
will be helpful in Chapter~\ref{secmix}, where by bringing the wave functions in different regions of
space-time ``out of phase'', we will be able to turn off causal influences between these regions.

\section{Preliminaries on Projectors, Fock Spaces and Entanglement} \label{sec2}
In this chapter we first recall the notions of the fermionic Fock space and entanglement, also fixing
our notation. Then we show that a projector in the one-particle Hilbert space uniquely determines
a Hartree-Fock state in the fermionic Fock space, making it impossible to describe entangled states.

\subsection{The Fermionic Fock Space and Entanglement} \label{sec21}
In non-relativistic quantum mechanics, the one-particle states form a separable Hilbert
space~$(\H, \la .|. \ra)$. Similarly, in Dirac theory we let~$\H$ be the Hilbert space corresponding
to the scalar product~\eqref{print}.
In each of these settings, a many-fermion state is
usually described by a vector in the fermionic Fock space, which we now introduce
(see also~\cite[Section~II.4]{reed+simon} or~\cite[Section~I.1]{schwabl2}).
We let~$\H^n = \H \otimes \cdots \otimes \H$ be the $n$-fold tensor product, endowed with
the natural scalar product
\beq \label{sdef}
\la \psi_1 \otimes \cdots \otimes \psi_n \,|\, \phi_1 \otimes \cdots \otimes \phi_n \ra 
:= \la \psi_1 | \phi_1 \ra\: \cdots\: \la \psi_n | \phi_n \ra\:.
\eeq
Totally anti-symmetrizing the tensor product gives the wedge product
\beq \label{wdef}
\psi_1 \wedge \cdots \wedge \psi_n := \frac{1}{n!} \sum_{\sigma \in S_n}
(-1)^{\sign(\sigma)}\: \psi_{\sigma(1)} \otimes \cdots \otimes \psi_{\sigma(n)}
\eeq
(here~$S_n$ denotes the set of all permutations and~$\sign(\sigma)$ is the sign of the permutation~$\sigma$).
The wedge product gives rise to a mapping
\[ \Lambda_n \::\: \underbrace{\H \times \ldots \times \H}_{\text{$n$ factors}} \rightarrow \H^n
\::\: (\psi_1, \ldots, \psi_n) \mapsto \psi_1 \wedge \cdots \wedge \psi_n \:. \]
We denote the image of this mapping by~$\FHF_n$. The vectors in~$\FHF_n$
are called $n$-particle {\em{Hartree-Fock states}} or {\em{factorizable states}}.
These states do in general not form a vector space, as the following
example shows, which in discussions of spin correlation experiments and Bell's inequalities
is often referred to as the EPR singlet state (see for example~\cite[Section~1.5]{afriat}).

\begin{Example} (The spatially separated singlet state) \label{exsinglet} {\em{We consider the
one-particle Hilbert space~$\H=\C^2_A \oplus \C^2_B$ of two spins (in quantum information
theory called ``qubits''), located at the positions of two
observers \textsc{Alice} and \textsc{Bob}. Choosing in~$\C^2$ the standard basis~$\psiu=(1,0)$
and~$\psid=(0,1)$ yields the basis~$(\psiu_A, \psid_A, \psiu_B, \psid_B)$ of~$\H$. The spatially separated singlet state is the following
linear combination of $2$-particle Hartree-Fock states
\beq \label{singlet}
\Psi := \frac{1}{\sqrt{2}} \left( \psiu_A \wedge \psid_B -
\psid_A \wedge \psiu_B \right) .
\eeq
Let us verify in detail that this state is not factorizable. Thus assume conversely that~$\Psi$
can be written as a product,
\[ \Psi = \psi_1 \wedge \psi_2 \:. \]
Computing this wedge product in the basis representations
\begin{align}
\psi_1 &= \alpha^\uparrow_A \psiu_A + \alpha^\downarrow_A \psid_A
+ \alpha^\uparrow_B \psiu_B +\alpha^\downarrow_B \psid_B \label{p1} \\
\psi_2 &= \beta^\uparrow_A \psiu_A + \beta^\downarrow_A \psid_A
+ \beta^\uparrow_B \psiu_B +\beta^\downarrow_B \psid_B\:,
\end{align}
the vanishing of the term~$\sim \psiu_A \wedge \psid_A$
implies that the vectors~$\alpha^\uparrow_A \psiu_A + \alpha^\downarrow_A \psid_A$
and~$\beta^\uparrow_A \psiu_A + \beta^\downarrow_A \psid_A$ must be
linearly dependent. Similarly, the vanishing of the term~$\sim \psiu_B \wedge \psid_B$
implies that the vectors~$\alpha^\uparrow_B \psiu_B +\alpha^\downarrow_B \psid_B$
and~$\beta^\uparrow_B \psiu_B +\beta^\downarrow_B \psid_B$ are linearly dependent.
Hence~$\psi_2$ can be written as
\beq \label{p22}
\psi_2 = \beta_A \left(\alpha^\uparrow_A \psiu_A + \alpha^\downarrow_A \psid_A \right)
+ \beta_B \left(\alpha^\uparrow_B \psiu_B +\alpha^\downarrow_B \psid_B \right)
\eeq
with suitable complex coefficients~$\beta_A$ and~$\beta_B$.
Taking the wedge product of~\eqref{p1} and~\eqref{p22} yields
\[ \Psi = \psi_1 \wedge \psi_2 = (\beta_B - \beta_A) \left(\alpha^\uparrow_A \psiu_A + \alpha^\downarrow_A \psid_A \right) \wedge \left(\alpha^\uparrow_B \psiu_B +\alpha^\downarrow_B \psid_B \right)\:. \]
Multiplying out and comparing with~\eqref{singlet}, one sees that the products~$\alpha^\uparrow_A \alpha^\downarrow_B$
and~$\alpha^\downarrow_A  \alpha^\uparrow_B$ must be non-zero, and
thus none of these four coefficients vanishes. But then the
term~$\sim \psiu_A \wedge \psiu_B$ is non-zero, a contradiction.
}} \QEDrem \end{Example} \noindent

We denote the vector space generated by the $n$-particle Hartree-Fock states by
\[ \F_n = \overline{ \bra \Lambda_n(\H^n) \ket } \:. \]
Their direct sum is the {\em{fermionic Fock space}},
\[ \F = \bigoplus_{n=0}^\infty \F_n \:. \]
The non-factorizable vectors~$\Psi \in \F_n \setminus \Lambda_n(\H^n)$ are called
{\em{entangled states}}. The spatially separated singlet state is the standard example
of an entangled state.
Entanglement is a basic phenomenon of quantum physics with
important potential applications in quantum computing.

\subsection{Projectors and Hartree-Fock States} \label{sechartree}
Let us examine in which sense a projector in the one-particle Hilbert space
characterizes a many-particle quantum state. Thus let~$P$ be a projector in the Hilbert
space~$(\H, \la .|. \ra)$, for simplicity of finite rank~$f$, i.e.
\beq \label{Pdef}
P^* = P = P^2 \qquad \text{and} \qquad \dim P(\H) = f \:.
\eeq
In order to get a connection to the fermionic Fock space
formalism, we choose an orthonormal basis~$\psi_1, \ldots, \psi_f$ of~$P(\H)$ and form
the Hartree-Fock state
\beq \label{wedge}
\Psi := \psi_1 \wedge \cdots \wedge \psi_f \in \FHF_f\:.
\eeq
The choice of our orthonormal basis was unique only up to the unitary transformations
\beq \label{btrans}
\psi_i \rightarrow \tilde{\psi}_i = \sum_{j=1}^f U_{ij} \psi_j \quad \text{with} \quad U \in \U(f)\:.
\eeq
Due to the anti-symmetrization, this transformation changes the corresponding Hartree-Fock state
only by a phase factor,
\beq \label{psiphase}
\tilde{\psi}_1 \wedge \cdots \wedge \tilde{\psi}_f = \det U \;
\psi_1 \wedge \cdots \wedge \psi_f \:.
\eeq
Thus we can indeed associate to the projector~$P$ a Hartree-Fock state, which is well-defined up
to a phase. As the phase of~$\Psi$ has no physical significance,
the physical system is described equivalently by a projector~$P_f$ on the many-particle
state~$\Psi$, i.e.\ in bra-/ket notation\footnote{Here the factor~$f!$ comes about because,
according to our conventions~\eqref{wdef} and~\eqref{sdef},
\begin{align*}
\la \psi_1 &\wedge \cdots \wedge \psi_f \:|\: \psi_1 \wedge \cdots \wedge \psi_f \ra
= \la \psi_1 \wedge \cdots \wedge \psi_f \:|\: \psi_1 \otimes \cdots \otimes \psi_f \ra \\
&= \frac{1}{f!} \sum_{\sigma \in S_f}
(-1)^{\sign(\sigma)} \:\la \psi_{\sigma(1)} | \psi_1 \ra \cdots \la \psi_{\sigma(f)} | \psi_f \ra
= \frac{1}{f!}\:.
\end{align*} }
\beq \label{Ppure}
P_f = \frac{1}{\|\Psi\|_\F^2}\: | \Psi \ra \la \Psi |
= f! \: |\psi_1 \wedge \cdots \wedge \psi_f \ra \la \psi_1 \wedge \cdots \wedge \psi_f|
\::\: \F_f \rightarrow \F_f \:.
\eeq
Since the phase freedom drops out when forming the projector~\eqref{Ppure},
this operator is well-defined. The next proposition gives an alternative definition of~$P_f$
which does not involve a choice of basis.
\begin{Prp} \label{prp13}
For any projector~$P$ in~$(\H, \la .|. \ra)$ of rank~$f$, the corresponding operator
\beq \label{Pndef}
P_f \::\: \F_f \rightarrow \F_f \::\: \psi_1 \wedge \cdots \wedge \psi_f
\rightarrow (P \psi_1) \wedge \cdots \wedge (P \psi_f)
\eeq
is a projector onto an $f$-particle Hartree-Fock state. The mapping $P \rightarrow P_f$
gives a one-to-one correspondence between projectors in~$\H$ and
projectors on Hartree-Fock states in~$\F$.
\end{Prp}
\Proof It follows immediately from the definitions that~$P_f$ is symmetric and idempotent,
and is thus a projector.
To compute the rank of~$P_f$, we choose an orthonormal basis~$\psi_1, \ldots, \psi_f$
of~$P(\H)$ and extend it to an orthonormal basis of~$\H$. As is obvious from~\eqref{Pndef}, the operator
$P_f$ applied to any wedge product of basis vectors vanishes unless all basis vectors
are elements of the set~$\{\psi_1, \ldots, \psi_f\}$. Hence the vector
$\Psi := \psi_1 \wedge \cdots \wedge \psi_f$ is a basis of the image of~$P_f$.
We conclude that~$P_f$ has indeed rank one and is thus a projector onto
the Hartree-Fock state~$\Psi$.

Now suppose conversely that~$P_f$ is a projector onto a Hartree-Fock state.
Representing this operator in the form~\eqref{Ppure}, we let~$P$ be the projector in~$\H$
on the subspace~$\bra \psi_1, \ldots, \psi_f \ket$. Then the operator~$P_f$ has the
representation~\eqref{Pndef}, concluding the proof.
\QED

\subsection{Projectors and Expectation Values} \label{sec24}
We now consider how expectation values of observables can be expressed in terms of
the projectors~$P$ and~$P_f$.
We begin with a {\em{one-particle observable}}~$\cO$, being a self-adjoint operator
in the one-particle Hilbert space~$\H$. By
\begin{equation}
\begin{split}
\cO^\F (\psi_1 \wedge \cdots \wedge \psi_n) :=
(\cO \psi_1) \wedge \cdots \wedge\psi_n&
+ \psi_1 \wedge (\cO \psi_2) \wedge \cdots \wedge \psi_n \\
&+ \cdots + \psi_1 \wedge \cdots \wedge \psi_{n-1} \wedge (\cO \psi_n)  \label{oneO} 
\end{split}
\end{equation}
we can define a corresponding operator~$\cO^\F$ on the Fock space~$\F$. This operator preserves the number of particles in the sense that it maps the $n$-particle subspace~$\F_n$ to itself.

Suppose that an $f$-fermion state is described by a projector~$P$, \eqref{Pdef}.
The next lemma shows how the expectation values of $\cO^\F$ and of
products of one-particle operators can be expressed in terms of traces involving the projector~$P$.
\begin{Lemma} \label{lemma24} Suppose that~$\cO$ and $\cO_{1\!/\!2}$ are one-particle
observables. Describing a many-fermion state by a projector~$P$ in~$(\H, \la .|. \ra)$, we have
\begin{align}
\la \cO^\F \ra &= \Tr_\H (P \cO ) \label{tr1} \\
\la \cO_1^\F \cO_2^\F \ra &= \Tr_\H (P \cO_1 \cO_2)
+ \Tr_\H (P \cO_1 ) \Tr_\H (P \cO_2) - \Tr_\H (P \cO_1 P \cO_2)\:.
\label{tr2}
\end{align}
\end{Lemma}
\Proof The expectation values are obtained by
taking the trace of the observables multiplied by the operator~$P_f$,
\[ \la \cO^\F \ra = \Tr_{\F_f}(P_f \cO ^\F) \:,\qquad
\la \cO_1^\F \cO_2^\F \ra = \Tr_{\F_f}(P_f \cO_1^\F \cO_2 ^\F) \:. \]
Representing~$P_f$ in the form~\eqref{Ppure}, it follows that
\begin{align*}
\la \cO^\F  \ra &= f! \: \la \psi_1 \wedge \cdots \wedge \psi_f
\:|\: \cO^\F \left( \psi_1 \wedge \cdots \wedge \psi_f \right) \ra \\
&= f! \: \la \psi_1 \wedge \cdots \wedge \psi_f
\:|\: \cO^\F \left( \psi_1 \otimes \cdots \otimes \psi_f \right) \ra \\
&= \sum_{\sigma \in S_n} \la \psi_{\sigma(1)} \otimes \cdots \otimes \psi_{\sigma(f)}
\:|\: \cO^\F \left( \psi_1 \otimes \cdots \otimes \psi_f \right) \ra 
= \sum_{i=1}^f \la \psi_i \:|\: \cO \,\psi_i \ra\:,
\end{align*}
where in the last step we applied~\eqref{oneO} together with~\eqref{sdef} and used the fact
that the vectors~$\psi_1, \ldots, \psi_f$ are orthonormal. This proves~\eqref{tr1}.

With the same method, we obtain
\begin{align*}
\la \cO_1^\F &\cO_2^\F \ra = f! \: \la \psi_1 \wedge \cdots \wedge \psi_f
\:|\: \cO_1^\F \cO_2^\F \left( \psi_1 \otimes \cdots \otimes \psi_f \right) \ra \\
=& \sum_{k=1}^f f! \: \la \psi_1 \wedge \cdots \wedge \psi_f
\:|\: \psi_1 \otimes \cdots \psi_{k-1} \otimes (\cO_1 \cO_2 \psi_k)
\otimes \psi_{k+1} \cdots \psi_f \ra \\
&+\sum_{k \neq l} f! \: \la \psi_1 \wedge \cdots \wedge \psi_f
\:|\: \psi_1 \otimes \cdots  (\cO_1 \psi_k) \cdots  (\cO_2 \psi_l) \cdots
\otimes \psi_f \ra \\
=& \sum_{k=1}^f \langle \psi_k | \cO_1 \cO_2 \psi_k \rangle
+ \sum_{k \neq l} \Big( \langle \psi_k | \cO_1 \psi_k \rangle
\langle \psi_l | \cO_2 \psi_l \rangle -
\langle \psi_k | \cO_1 \psi_l \rangle \langle \psi_l | \cO_2 \psi_k \rangle \Big) .
\end{align*}
The last sum can be extended to all~$k,l=1,\ldots, f$, because the summands for~$k=l$
vanish. We thus obtain~\eqref{tr2}.
\QED
The method of this lemma immediately extends to higher powers of one-particle observables.

More generally, one can consider {\em{many-particle observables}}, described
by a self-adjoint operator~$\cO$ on the Fock space~$\F$. In this paper, we shall only consider
observables which {\em{preserve the number of particles}},
i.e.\ which are invariant on the $n$-particle subspaces~$\F_n$,
\beq \label{Fninv}
\cO : \F_n \rightarrow \F_n\:.
\eeq
This assumption is justified by the physical law of the conservation of the baryon and lepton numbers,
stating that the total number of fermions is preserved.
Thus by considering a system which is so large that no fermion enters or leaves it,
we can arrange that all physical observables satisfy~\eqref{Fninv}.
For a many-particle observable satisfying~\eqref{Fninv}, the expectation value is again expressed by
a trace,
\beq \label{expect}
\la \cO  \ra = \Tr_{\F_f} \!\left( P_f \cO \right)\:.
\eeq

\section{Microscopic Mixing of  Decoherent Space-Time Regions} \label{secmix}
In this chapter we develop a method for describing entangled states by a projector in
the one-particle Hilbert space. In Section~\ref{secmm} we give a preliminary construction,
which clarifies the difficulty in describing entangled states.
In Section~\ref{secindepend} we overcome this difficulty on a rather
formal and axiomatic level.
The microscopic justification of the resulting formalism will be given in
Sections~\ref{secmmrs}--\ref{secFsuper} in the framework of the fermionic projector.
% Nehme nächste zwei Sätze für entangleshort heraus!
Appendix~\ref{appA1} explains an attempt to describe entanglement by restricting
attention to a subsystem. Seeing why this attempt fails might serve to the reader
as a motivation for the concept of microscopic mixing.

\subsection{Microscopic Mixing of the Wave Functions} \label{secmm}
\begin{figure}
\begin{center}
\begin{picture}(0,0)%
\includegraphics{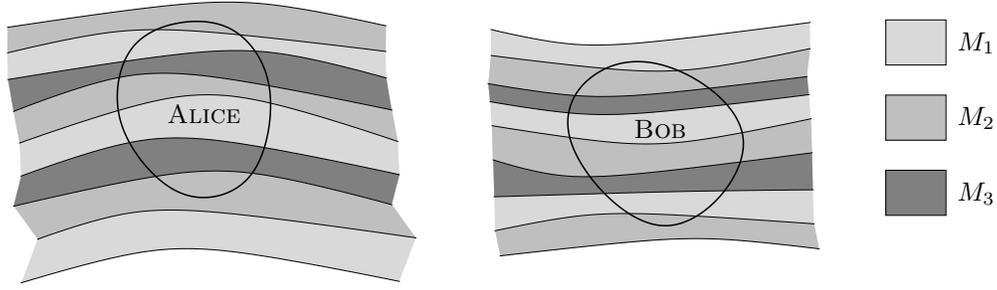}%
\end{picture}%
\setlength{\unitlength}{829sp}%
\begingroup\makeatletter\ifx\SetFigFontNFSS\undefined%
\gdef\SetFigFontNFSS#1#2#3#4#5{%
  \reset@font\fontsize{#1}{#2pt}%
  \fontfamily{#3}\fontseries{#4}\fontshape{#5}%
  \selectfont}%
\fi\endgroup%
\begin{picture}(29017,8452)(-1551,-308)
\put(27451,6689){\makebox(0,0)[b]{\smash{{\SetFigFontNFSS{10}{12.0}{\rmdefault}{\mddefault}{\updefault}$M_1$}}}}
\put(27451,4439){\makebox(0,0)[b]{\smash{{\SetFigFontNFSS{10}{12.0}{\rmdefault}{\mddefault}{\updefault}$M_2$}}}}
\put(27451,2189){\makebox(0,0)[b]{\smash{{\SetFigFontNFSS{10}{12.0}{\rmdefault}{\mddefault}{\updefault}$M_3$}}}}
\put(18001,4049){\makebox(0,0)[b]{\smash{{\SetFigFontNFSS{10}{12.0}{\rmdefault}{\mddefault}{\updefault}$\text{\textsc{Bob}}$}}}}
\put(4351,4499){\makebox(0,0)[b]{\smash{{\SetFigFontNFSS{10}{12.0}{\rmdefault}{\mddefault}{\updefault}$\text{\textsc{Alice}}$}}}}
\end{picture}%
\caption{Example of microscopic mixing in a spin correlation experiment.}
\label{fig1}
\end{center}
\end{figure}

Our idea for realizing entangled states is to give~$P$
a {\em{non-trivial microscopic structure}}, with the hope that ``averaging'' this microstructure
over macroscopic regions of space-time will give rise to an effective
kernel~$P(x,y)$ of a more general form which allows for the description of entanglement.
As the relativistic generalization will not be quite straightforward, we begin for clarity in the
non-relativistic setting. Thus we consider the situation where space is subdivided into
sets $M_1, \ldots, M_L$,
\beq \label{Msep}
M = M_1 \cup \cdots \cup M_L\qquad \text{and} \qquad M_a \cap M_b = \varnothing \quad
\text{if $a \neq b$},
\eeq
which are fine-grained in the sense that every
macroscopic region of space-time intersects several of the sets~$M_a$. The sets $M_a$
may be localized, but they can also be extended over a macroscopic region of space-time,
for example by forming ``layers'' or ``filaments'' connecting the two observers in the spin correlation
experiment of Example~\ref{exsinglet} (see Figure~\ref{fig1}).
The macroscopic physical objects are then introduced by homogenizing over the sets~$M_a$.
We refer to this technique as the {\em{method of microscopic mixing}}.

The partition~\eqref{Msep} allows us to decompose~$\H$ 
into an orthogonal direct sum of the Hilbert spaces~$\H_a$ of square integrable
wave functions in~$M_a$,
\beq \label{Hdec}
\H = \bigoplus_{a=1}^L \H_a\:.
\eeq
Thus every wave function~$\psi_i$ in the image of~$P$, \eqref{wedge}, has the unique
decomposition
\beq \label{psisep}
\psi_i = \sum_{a=1}^L \psi_i^{(a)} \qquad \text{with} \qquad \psi_i^{(a)} \in \H_a\:.
\eeq
The simplest attempt is to realize a macroscopic local one-particle observable~${\mathcal{O}}$ (like
a position or spin operator) by an operator on~$\H$ being invariant on~$\H_a$,
\beq \label{ainv}
{\mathcal{O}} \,:\, \H_a \rightarrow \H_a\:.
\eeq
Then the corresponding one-particle expectation values split into a sum over the
subsystems,
\[ \la \psi | {\mathcal{O}} \psi \ra = \sum_{a=1}^L \la \psi^{(a)} | {\mathcal{O}} \psi^{(a)}
\ra = \sum_{a=1}^L \int_{M_a} \overline{\psi}(x)
{\mathcal{O}} \psi(x)\: dx \:, \]
and this can be understood as an ``averaging process'' over the subregions~$M_a$.

Following the constructions in Section~\ref{sec24}, every one-particle operator induces
a corresponding operator on the Fock space~$\F$, and products of such operators
yield corresponding many-particle observables. Taking expectation values of such operators
in the Fock space again involves an ``averaging process'' over the subregions~$M_a$.
More specifically, describing the many-particle system by a projector~$P$ in~$\H$,
the expectation value of the two-particle observables corresponding
to \textsc{Alice} and \textsc{Bob} is given by
(see Lemma~\ref{lemma24})
\begin{align}
\la \cO_A^\F \cO_B^\F \ra &=  \sum_{i=1}^f \sum_{a=1}^L
\la \psi_i^{(a)} | \cO_A \cO_B \psi_i^{(a)} \ra \label{expsimple0} \\
+ \sum_{i,j=1}^f & \sum_{a,b=1}^L \Big(
\la \psi_i^{(a)} | \cO_A \psi_i^{(a)} \ra \la \psi_j^{(b)} | \cO_B \psi_j^{(b)} \ra
- \la \psi_i^{(a)} | \cO_A \psi_j^{(a)} \ra \la \psi_j^{(b)} | \cO_B \psi_i^{(b)} \ra \Big) .
\label{expsimple}
\end{align}
In~\eqref{expsimple}, an ``averaging process'' takes place at each of the observers.
However, it is important to note that there is no averaging over {\em{correlations}} between
the two observers, as would be the case for an expression like
\beq \label{avcor}
\sum_{i,j=1}^f \sum_{a=1}^L 
\la \psi_i^{(a)} | \cO_A \psi_i^{(a)} \ra \la \psi_j^{(a)} | \cO_B \psi_j^{(a)} \ra \:.
\eeq
% We conclude that the above method does {\em{not}} make it possible
% to realize general entangled states.
% Nehme die nächsten beiden Sätze für entangleshort heraus; dafür dann der Satz davor!
This shortcoming is the basic reason why the above method does {\em{not}} make it possible
to realize general entangled states, as is worked out in Appendix~\ref{appA2}.
Thus our task is to extend our framework so as to
obtain averages over correlations~\eqref{avcor}.

\subsection{A Formalism for the Description of Entanglement} \label{secindepend}
The previous construction did not take into account
that the measurement process is a result of an interaction of the system with the measurement
device. Assuming that the subsystems have an {\em{independent dynamics}} (an assumption
which will be justified in Section~\ref{secdecind} below), also the measurement
process should take place independently in the subsystems. Following this idea makes it
possible to describe entanglement, as we now explain.

For the one-particle observables, the assumption of an independent dynamics of
the subsystems was already taken into account in~\eqref{ainv} by the assumption that~${\mathcal{O}}$
should be invariant on the subspaces~$\H_a$. However, for a many-particle observable, it was too
naive to simply take the product of the one-particle operators
(see~\eqref{expsimple0} and~\eqref{expsimple}). Thinking of the situation in Figure~\ref{fig1},
\textsc{Alice} is built up of fermionic wave functions. Thus considering her as part of the physical system,
we should replace the corresponding measurement operator~${\mathcal{O}}_A$ by separate
operators~${\mathcal{O}}_A^{(a)}$ for each of the subsystems. Proceeding similarly for \textsc{Bob},
in the subsystem~$M_a$ measurements are to be carried out with the operators~${\mathcal{O}}_A^{(a)}$
and~${\mathcal{O}}_B^{(a)}$. For a correlation measurement in~$M_a$, we should
extend the one-particle observables~${\mathcal{O}}^{(a)}$
to operators~${\mathcal{O}}^{\F^{(a)}}$ defined on the Fock space~$\F^{(a)}_n$ of the subsystem
given by
\beq \label{Fockn}
\F^{(a)}_n = \overline{\bra \Lambda_n(\H_a^n) \ket} \subset \F_n
\eeq
(as explained after~\eqref{Fninv}, we again restrict attention to observables which preserve the number
of particles) and consider the corresponding two-particle observable
\[ {\mathcal{O}}_A^{\F^{(a)}} {\mathcal{O}}_B^{\F^{(a)}} \::\: \F^{(a)}_n \rightarrow
\F^{(a)}_n\:. \]
More generally, we make the following assumption:
\begin{itemize}
\item[(A)] The observables correspond to operators~${\mathcal{O}}$ which are invariant
on~$\F^{(a)}_n$,
\[ {\mathcal{O}} : \F^{(a)}_n \rightarrow \F^{(a)}_n \:,\qquad a=1,\ldots, L\:. \]
\end{itemize}
Similar as explained in Section~\ref{sechartree} for the Fock space~$\F_f$,
we can get a simple connection between a projector~$P$ in~$\H$ and the
Fock spaces~$\F^{(a)}_f$. Namely, choosing again an orthonormal basis~$\psi_1, \ldots, \psi_f$
of~$P(\H)$ and decomposing
each of the one-particle wave functions according to~\eqref{psisep}, we can construct
Hartree-Fock states~$\Psi^{(a)}$ in the $f$-particle Fock spaces of the subsystems,
\beq \label{Psiadef}
\Psi^{(a)} := \psi_1^{(a)} \wedge \cdots \wedge \psi_f^{(a)} \in \F_f^{(a)}\:.
\eeq
Exactly as in~\eqref{psiphase}, one sees that these vectors are unique up
to a common phase,
\beq \label{comphase}
\Psi^{(a)} \rightarrow e^{i \varphi}\: \Psi^{(a)} \qquad \text{with} \qquad \varphi \in \R
\text{ independent of~$a$}\:.
\eeq

The setting so far is not sufficient for determining the expectation value of a measurement,
because for computing an expectation value we need to take an ``average'' over the subsystems.
This process can be described conveniently by the so-called {\em{measurement scalar
product}}~$(.|.)$, which we now introduce
(for a microscopic derivation of the measurement scalar product
and the measurement process see Section~\ref{secFsuper} below).
It is defined on the one-particle Hilbert space~$(\H, \la .|. \ra)$ as
a positive semi-definite inner product
\beq \label{sprodm}
(.|.) \::\: \H \times \H \rightarrow \C\:,
\eeq
with respect to which the direct sum decomposition~\eqref{Hdec} need not be orthogonal.
The fact that this inner product is only semi-definite models the fact that the measurement
process may involve a homogenization process on the microscopic scale, so that
fluctuations of the wave functions on the small scale might not enter the measurement process.
The measurement scalar product induces on the Fock spaces the bilinear form
\[ \begin{split}
(.|.)^{(a,b)}\::\: \F^{(a)}_n \times \F^{(b)}_n \rightarrow \C \::\: (&\psi_1^{(a)} \wedge \cdots \wedge \psi_n^{(a)},\: \psi_1^{(b)} \wedge \cdots \wedge \psi_n^{(b)}) \\
&\mapsto \frac{1}{n!} \sum_{\sigma \in S_n} (-1)^{\sign(\sigma)}
(\psi^{(a)}_{\sigma(1)} | \psi^{(b)}_1) \cdots (\psi^{(a)}_{\sigma(n)} | \psi^{(b)}_n) \:.
\end{split} \]
We now specify how expectation values are to be computed and state the assumptions
which ensure that these expectation values are real.
\begin{itemize}
\item[(B)] The expectation value of the measurement of the observable~${\mathcal{O}}$
is given by
\beq \label{Bsum}
\la {\mathcal{O}} \ra = \frac{\sum_{a,b=1}^L ( \Psi^{(a)} \,|\, {\mathcal{O}}\, \Psi^{(b)})^{(a,b)} }
{\sum_{a,b=1}^L ( \Psi^{(a)} \,|\, \Psi^{(b)})^{(a,b)} } \:.
\eeq
\item[(C)] The observables are symmetric possibly up to a microscopic error, meaning that
\[ ( \Psi^{(a)} \,|\, {\mathcal{O}} \Psi^{(b)})_{\F} =  ( {\mathcal{O}} \Psi^{(a)} \,|\, \Psi^{(b)})_{\F}
+ \O(\varepsilon) \:, \]
where~$\varepsilon$ is the length scale of microscopic mixing.
\end{itemize}
Finally, we need to specify how a state changes in a measurement process. 
In order to ensure that a repeated measurement of the same observable yields identical
results, one usually asserts that after the measurement, the state should be an eigenstate of
the observable.
In our setting, the situation is a bit more involved because the measurement
process may change the number of subsystems, and only the wave function after the homogenization
should be an eigenstate of the observable. This is made precise by the following construction.
We take the direct sum of the vector spaces~$\F_n^{(a)}$,
\[ {\mathscr{G}} := \bigoplus_{a=1}^L \F_n^{(a)} \:, \]
and on these spaces we introduce the inner product
\[ (.|.)_{\mathscr{G}} : {\mathscr{G}} \times {\mathscr{G}} \rightarrow \C \::\:
\Big((\Psi^{(a)})_{a=1,\ldots, L}, (\Psi^{(b)})_{a=1,\ldots, L} \Big) \mapsto
\sum_{a,b=1}^L (\Psi^{(a)} | \Psi^{(b)})^{(a,b)}\:. \]
This inner product is positive semi-definite, but it need not be definite.
Dividing out the null space and taking the completion,
\beq \label{Feffdef}
\F^\text{eff}_n := \overline{{\mathscr{G}} / {\mathscr{G}}_0} \qquad \text{where} \qquad
{\mathscr{G}}_0 = \{ u \in {\mathscr{G}} \text{ with } (u|u)_{\mathscr{G}}=0 \}\:,
\eeq
we obtain a Hilbert space, which we can regard as the effective $n$-particle Fock space
obtained by homogenization over the subsystems. We denote the natural projection operator by~$\pi_n$,
\beq \label{pindef}
\pi_n \::\: \bigoplus_{a=1}^L \F_n^{(a)} \rightarrow \F^\text{eff}_n\:.
\eeq
Using linearity together with Assumption~(C) above, every observable~${\mathcal{O}}$ induces an operator
\beq \label{Oeff}
{\mathcal{O}}^\text{eff} \::\: \F^\text{eff}_n \rightarrow \F^\text{eff}_n\:,
\eeq
uniquely defined possibly up to a microscopic error.
To the projector~$P$ in~$\H$ we associate a corresponding state
\beq \label{Psieff}
\Psi^\text{eff} = \pi_f\! \left( \Psi^{(1)}, \ldots, \Psi^{(L)} \right) \in \F_f^\text{eff}
\eeq
(with the~$\Psi^{(a)}$ as in~\eqref{Psiadef}). According to~\eqref{comphase}, this state
is well-defined up to an irrelevant phase.
\begin{itemize}
\item[(D)] After a measurement of the observable~${\mathcal{O}}$, the one-particle
projector~$P$ takes such a form that the corresponding state~$\Psi^\text{eff} \in \F_f^\text{eff}$
defined by~\eqref{Psieff} is an eigenstate of the operator~${\mathcal{O}}^\text{eff}$,
\eqref{Oeff}.
\end{itemize}
Similar as in the Copenhagen interpretation or the formulation of the measurement process
by von Neumann~\cite[Section~V.1]{vneumann},
the above Assumptions~(A)--(D) are merely working rules to determine the results of measurements.
For a conceptually convincing treatment, these assumptions
should be derived from the physical equations.

We now verify that the above setting indeed makes it possible to realize the EPR singlet state.
\begin{Example} (The spatially separated singlet state) \label{exssss} {\em{
We choose a microscopic length scale~$\varepsilon>0$ and subdivide position space~$M=\R^3$
into two subregions $M_1$ and~$M_2$ which form layers of width~$\varepsilon$,
\[ M_1 = \{ \vec{x} \in \R^3 \text{ with } [x_1/\varepsilon] \in 2 \Z \} \:,\qquad
M_2 = \{ \vec{x} \in \R^3 \text{ with } [x_1/\varepsilon] \in 2 \Z+1 \} \]
(where~$[x]=\min\{n \in \Z \:|\: n \geq x\}$ is the Gauss bracket). We introduce the wave functions
\begin{align*}
\psi_1(\vec{x}) &= \psiu_A(\vec{x})\: \chi_{M_1}(\vec{x}) + \psid_A(\vec{x})\: \chi_{M_2}(\vec{x}) \\
\psi_2(\vec{x}) &= \psid_B(\vec{x})\: \chi_{M_1}(\vec{x}) - \psiu_B(\vec{x})\: \chi_{M_2}(\vec{x}) \:,
\end{align*}
where~$\psi^{\uparrow/\downarrow}_{A/B}$ are smooth one-particle wave functions
supported near \textsc{Alice} or \textsc{Bob}
(and where~$\chi$ is the characteristic function defined by~$\chi_N(x)=1$ if~$x \in N$
and~$\chi_N(x)=0$ otherwise). Defining~$P$ as the projector on the subspace spanned by~$\psi_1$
and~$\psi_2$, the corresponding two-particle wave functions of the subsystems are
\[ \Psi^{(1)} = c \left( \psiu_A \wedge \psid_B \right) \chi_{M_1 \times M_1}
\qquad \text{and} \qquad
\Psi^{(2)} = -c \left( \psid_A \wedge \psiu_B \right)  \chi_{M_2 \times M_2} \]
with~$c$ a normalization constant.

In order to realize a suitable mixing of the subregions in the measurement process, we introduce
the measurement scalar product by
\[ ( \psi | \phi) = \int_M \overline{\psi(\vec{x})} \phi(\vec{x}) \,d\vec{x}
+ \frac{1}{2} \int_M \left( \overline{\psi(\vec{x}+ \varepsilon e_1)} \phi(\vec{x}) 
+ \overline{\psi(\vec{x})} \phi(\vec{x}+ \varepsilon e_1)  \right) d\vec{x} \:. \]
The spin operators are symmetric with respect to this inner product, whereas the position
operators are symmetric up to an error of order~$\varepsilon$,
\[ (\vec{x} \psi | \phi)  - (\psi | \vec{x} \phi) = \frac{\varepsilon}{2}
\int_M x_1 \left( \overline{\psi(\vec{x}+ \varepsilon e_1)} \phi(\vec{x}) 
- \overline{\psi(\vec{x})} \phi(\vec{x}+ \varepsilon e_1)  \right) d\vec{x}\:. \]
Thus the general observables introduced according to~(A) indeed satisfy the condition~(C).
The expectation values of the spin operators can now be calculated by applying the rule~(B).
More precisely, the inner products in~\eqref{Bsum} are computed by
\begin{align*}
(\Psi^{(1)} | {\mathcal{O}} \Psi^{(1)})^{(1,1)} &=
\la \Psi^{(1)} | {\mathcal{O}} \Psi^{(1)} \ra \\
(\Psi^{(2)} | {\mathcal{O}} \Psi^{(2)})^{(2,2)} &=
\la \Psi^{(2)} | {\mathcal{O}} \Psi^{(2)} \ra \\
(\Psi^{(1)} | {\mathcal{O}} \Psi^{(2)})^{(1,2)} &=
\frac{1}{2} \left( 
\la \Psi^{(1)}_+ | {\mathcal{O}} \Psi^{(2)} \ra
+ \la \Psi^{(1)} | {\mathcal{O}} \Psi^{(2)}_+ \ra \right) \\
(\Psi^{(2)} | {\mathcal{O}} \Psi^{(1)})^{(2,1)} &=
\frac{1}{2} \left( 
\la \Psi^{(2)}_+ | {\mathcal{O}} \Psi^{(1)} \ra
+ \la \Psi^{(2)} | {\mathcal{O}} \Psi^{(1)}_+ \ra \right) ,
\end{align*}
where on the right the scalar product on~$\F$ as defined by~\eqref{sdef} appears,
and the subscript~$+$ denotes that both spatial arguments of the corresponding two-particle
wave function have been shifted by~$\varepsilon e_1$.
Note that all these inner products involve integrals over~$M_1 \times M_1$
or~$M_2 \times M_2$. Since the wave functions~$\psi^{\uparrow\downarrow}_{A/B}$
are all smooth, we can extend the two-particle wave functions of the subsystems to
smooth functions on~$M \times M$,
\beq \label{Psi12def}
\Psi^{(1)}_\text{eff} = c \left( \psiu_A \wedge \psid_B \right)
\qquad \text{and} \qquad
\Psi^{(2)}_\text{eff} = c \left( \psid_A \wedge \psiu_B \right) .
\eeq
Shifting the arguments changes these smooth wave functions only by a term of order~$\varepsilon$.
Also, extending the integration range in the above integrals from~$M_1 \times M_1$
or~$M_2 \times M_2$ to $M \times M$ changes the values of the integrals only by a factor
of four, again up to contributions of order~$\varepsilon$. We thus obtain
\[ \sum_{a,b=1}^L ( \Psi^{(a)} \,|\, {\mathcal{O}}\,\Psi^{(b)})^{(a,b)}
= 4\: \Big\la (\Psi^{(1)}_\text{eff}+\Psi^{(2)}_\text{eff}) \,|\, {\mathcal{O}}\,
(\Psi^{(1)}_\text{eff}+\Psi^{(2)}_\text{eff})  \Big\ra + \O(\varepsilon) \:. \]
We conclude that in the limit~$\varepsilon \searrow 0$, the expectation values as defined by~(B)
indeed coincide with the expectation values of the spin singlet state.

Moreover, it is straightforward to verify that the space~$\F^\text{eff}_2$
as defined by~\eqref{Feffdef} can be identified with the ordinary Fock space~$\F_2$,
and that under this identification, the state~$\Psi^\text{eff}$ as given by~\eqref{Psieff}
goes over to the state~$\Psi^{(1)}_\text{eff}+\Psi^{(2)}_\text{eff}$ with
the wave functions~$\Psi^{(1 \!/\! 2)}_\text{eff}$ as in~\eqref{Psi12def}.
\QEDrem }}
\end{Example}

\subsection{Microscopic Mixing in the Framework of the Fermionic Projector} \label{secmmrs}
\begin{figure}
\begin{center}
\begin{picture}(0,0)%
\includegraphics{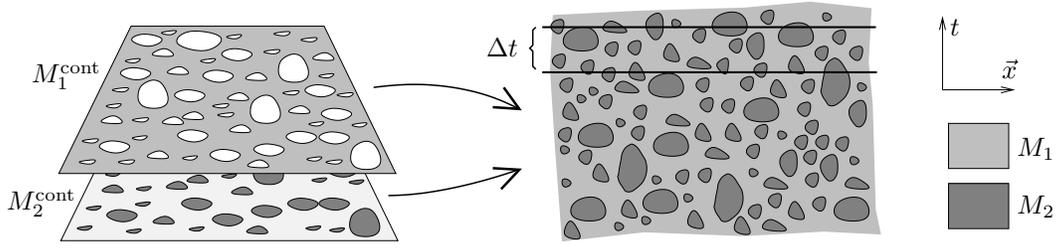}%
\end{picture}%
\setlength{\unitlength}{829sp}%
\begingroup\makeatletter\ifx\SetFigFontNFSS\undefined%
\gdef\SetFigFontNFSS#1#2#3#4#5{%
  \reset@font\fontsize{#1}{#2pt}%
  \fontfamily{#3}\fontseries{#4}\fontshape{#5}%
  \selectfont}%
\fi\endgroup%
\begin{picture}(29775,7182)(-1319,903)
\put(-554,5654){\makebox(0,0)[b]{\smash{{\SetFigFontNFSS{10}{12.0}{\rmdefault}{\mddefault}{\updefault}$M^\text{cont}_1$}}}}
\put(-1304,1844){\makebox(0,0)[b]{\smash{{\SetFigFontNFSS{10}{12.0}{\rmdefault}{\mddefault}{\updefault}$M^\text{cont}_2$}}}}
\put(12481,6449){\makebox(0,0)[b]{\smash{{\SetFigFontNFSS{10}{12.0}{\rmdefault}{\mddefault}{\updefault}$\Delta t$}}}}
\put(28441,3449){\makebox(0,0)[b]{\smash{{\SetFigFontNFSS{10}{12.0}{\rmdefault}{\mddefault}{\updefault}$M_1$}}}}
\put(28441,1739){\makebox(0,0)[b]{\smash{{\SetFigFontNFSS{10}{12.0}{\rmdefault}{\mddefault}{\updefault}$M_2$}}}}
\put(27646,5774){\makebox(0,0)[b]{\smash{{\SetFigFontNFSS{10}{12.0}{\rmdefault}{\mddefault}{\updefault}$\vec{x}$}}}}
\put(26026,7064){\makebox(0,0)[b]{\smash{{\SetFigFontNFSS{10}{12.0}{\rmdefault}{\mddefault}{\updefault}$t$}}}}
\end{picture}%
\caption{Example of a microscopic mixing of two space-time regions.}
\label{fig2}
\end{center}
\end{figure}

We now generalize and adapt the method of microscopic mixing to the framework
of the fermionic projector. For the extension to the relativistic setting, we decompose Minkowski
space~$M$ into two disjoint subregions
\beq \label{subs}
M = M_1 \cup M_2 \qquad \text{with} \qquad M_1 \cap M_2 = \varnothing \:,
\eeq
which may be fine-grained as depicted in Figure~\ref{fig2}
(for simplicity, we only consider two subregions; the generalization
to a finite number of subsystems is straightforward).
If one prefers, one can in addition replace the space-time continuum by a discrete set of points,
but the distinction between continuum and discrete space-time will not be of relevance for
the following considerations. We again consider a family
of wave functions~$\psi_1, \ldots, \psi_f$, where in view of the fact that we
also count the states of the Dirac sea, the number~$f$ of particles is very large
(see Section~\ref{seccontinuum}). As in~\eqref{Pdiscrete}, the fermionic projector
takes the form
\beq \label{Pxydef}
P(x,y) = -\sum_{j=1}^f |\psi_j(x) \Sr \Sl \psi_j(y) | \qquad \text{with} \qquad
x,y \in M_1 \cup M_2\:.
\eeq
Splitting up the wave functions similar to~\eqref{psisep} by
\beq \label{Psplit}
\psi_j = \psi_j^{(1)} + \psi_j^{(2)} \qquad \text{with} \qquad
\psi_j^{(a)} = \psi_j \: \chi_{M_a} \quad (a=1,2)\:,
\eeq
to each subsystem we can associate similar to~\eqref{Psiadef} the many-particle wave function
\beq \label{Psisub}
\Psi^{(a)} = \psi_1^{(a)} \wedge \cdots \wedge \psi_f^{(a)}  \:.
\eeq

At this point, we need to discuss the significance of the inner products~\eqref{stip} and \eqref{print}.
In discrete space-time, the underlying inner product involves the sum over all space-time
points~\eqref{stipdiscrete}. Likewise, in the
continuum limit, the sum is to be replaced by a space-time integral~\eqref{stip}.
The notions of symmetry and idempotence of the fermionic projector refer to this indefinite
inner product (for details see~\cite[\S2.6]{PFP}). However, as the fermionic
projector is built up of negative definite states, we may always restrict attention to
a negative definite subspace, on which~$\la .|. \ra := -\bra .|. \ket$ is a scalar product.
After completion, we again get a separable Hilbert space~$(\H, \la .|. \ra)$, so that we are back in
the setting of Chapter~\ref{sec2}.
It is important to keep in mind that the inner product~\eqref{stip} is not suitable for computing the expectation value of a measurement, which usually takes place
at a fixed time. Thus in the measurement process, it is natural to work
with a different scalar product, which can be regarded as the relativistic analog of the
{\em{measurement scalar product}}~\eqref{sprodm}. In view
of the fact that the integrand of~\eqref{print} has the interpretation as the probability density,
the scalar product~\eqref{print} seems the correct choice.

Let us consider how to implement the scalar product~\eqref{print} as the measurement scalar product
in the presence of a microscopic mixing of two subsystems.
First, taking into account that realistic measurements take place in a finite time interval, it seems a
good idea to replace the spatial integral in~\eqref{subs} by an integral over a strip of width~$\Delta t$
(as shown in Figure~\ref{fig2}). Moreover, the measurement should involve some kind of
homogenization process on the microscopic scale.
This is clear empirically because the expectation value must be a computable quantity
which involves taking averages over the subsystems. The homogenization
can also be understood microscopically from the fact that the measurement devices
are themselves formed of quantum mechanical wave functions which are spread out in space-time.
A typical example for an idealized measurement device is the
operator~$|\eta \ket \!\!\bra \eta|$, where~$\eta$ is a wave function which is supported inside
a strip of width~$\Delta t$ as shown in Figure~\ref{fig2}. The simplest way to take into account
the effect of such a measurement device would be to consider instead of~\eqref{print} the
measurement scalar product
\beq \label{measex}
( \psi | \phi ) = \frac{1}{(\Delta t)^2} \int_{\R^3} d\vec{x}
\int_{t}^{t+\Delta t} dt_1 \int_{t}^{t+\Delta t} dt_2\:
\Sl \psi(t_1, \vec{x}) | \gamma^0 \phi(t_2,\vec{x}) \Sr \:,
\eeq
where we take the average over a small time interval before computing the spatial inner product.
A more realistic measurement device is of course much more complicated, but
fortunately the details are of no relevance. All that matters is that the measurement scalar product
involves a homogenization process, with the effect that the direct summands in~\eqref{Hdec}
are in general not orthogonal with respect to~$(.|.)$ and that the null space~${\mathscr{G}}_0$
in~\eqref{Feffdef} becomes non-trivial.

\subsection{Justification of an Independent Dynamics of the Subsystems} \label{secdecind}
Following the concept of discrete space-time of Section~\ref{secdiscrete}, space-time
is not smooth on the microscopic scale, but should have a non-trivial microstructure.
In order to explain the possible consequences of such a microstructure in a simple setting,
we consider what happens if we choose the wave functions~$\psi_j^{(a)}$ in the two
subregions differently.  More specifically, we transform the wave functions in the second subsystem 
by a unitary matrix of determinant one,
\beq \label{decoher}
\psi_j^{(1)} \rightarrow \psi_j^{(1)} \:,\qquad
\psi_j^{(2)} \rightarrow \tilde{\psi}_j^{(2)} :=
\sum_{k=1}^f U_{jk} \,\psi_k^{(2)} \quad \text{with} \qquad U \in \SU(f)\:.
\eeq
This transformation has the advantage that it has no effect on the many-particle wave
functions~\eqref{Psisub}, because (exactly as in~\eqref{psiphase})
\beq \label{Psi2}
\Psi^{(2)} \rightarrow \tilde{\Psi}^{(2)} = \tilde{\psi}_1^{(2)} \wedge \cdots \wedge \tilde{\psi}_f^{(2)}
= \det U \; \psi_1^{(2)} \wedge \cdots \wedge \psi_f^{(2)} = \Psi^{(2)} \:.
\eeq
Furthermore, the fermionic projector does not change if its two arguments are
in the same subsystem, because in the case~$x,y \in M_2$, the unitarity of~$U$ yields that
\begin{align*}
P(x,y) \rightarrow &-\sum_{j=1}^f  |\tilde{\psi}^{(2)}_j(x) \Sr \Sl \tilde{\psi}^{(2)}_j(y) |
= -\sum_{j,k=1}^f (U U^\dagger)_{jk} \:| \psi^{(2)}_j(x) \Sr \Sl \psi^{(2)}_k(y) | \\
&= -\sum_{j}^f |\psi^{(2)}_j(x) \Sr \Sl \psi^{(2)}_j(y) | = P(x,y) \:.
\end{align*}
However, if the two arguments of the fermionic projector are in different space-time regions,
the operator~$U$ does not drop out. For example, if~$x \in M_2$ and~$y \in M_1$,
\beq \label{Psum}
P(x,y) \rightarrow -\sum_{j=1}^f |\tilde{\psi}^{(2)}_j(x) \Sr \Sl \psi^{(1)}_j(y) |
= -\sum_{j,k=1}^f U_{jk} \:| \psi^{(2)}_j(x) \Sr \Sl \psi^{(1)}_k(y) | \:.
\eeq
In the special case when~$U$ is a diagonal matrix whose entries are phase factors,
\[ U = \text{diag}(e^{i \varphi_1}, \ldots, e^{i \varphi_f}) \qquad \text{with} \qquad
\sum_{j=1}^f \varphi_j = 0 \mod 2 \pi \:, \]
the summations in~\eqref{Psum} reduce to one sum involving the phase factors,
\[ P(x,y) \rightarrow -\sum_{j=1}^f e^{i \varphi_j} \:|\psi^{(2)}_j(x) \Sr \Sl \psi^{(1)}_j(y) | \:. \]
If the angles~$\varphi_j$ are chosen stochastically, the phases of the summands are random.
As a consequence, there will be cancellations in the sum, and keeping in mind that the
number of summands is very large, we conclude that~$P(x,y)$ will be very small.
More generally, we find that if~$U$ is a random matrix,
$P(x,y)$ becomes small if~$x$ and~$y$ lie in different subsystems
(this argument will be quantified in Section~\ref{secrandom} below by integrating over the
space of unitary matrices).

From the physical point of view, the above consideration can be understood
using the notion of decoherence. If the one-particle wave functions~$\psi_j^{(1)}$
and~$\psi_j^{(2)}$ are coherent or ``in phase'', then the fermionic projector~$P(x,y)$
has the usual form, no matter whether~$x$ and~$y$ are in the same subsystem or not.
If however the wave functions in the subregions are decoherent or ``out of phase'', then the
fermionic projector~$P(x,y)$ will be very small if~$x$ and~$y$ are in different subregion.
We refer to this effect as the {\em{decoherence between space-time regions}}.
It should be carefully distinguished from the decoherence of the many-particle wave function
(see for example~\cite{joos}). Namely, as we saw in~\eqref{Psi2}, in our case
the many-particle wave functions remain unchanged. Thus they remain coherent, no
decoherence in the sense of~\cite{joos} appears. But the one-particle wave functions
become decoherent~\eqref{decoher}, having an influence on the fermionic projector~\eqref{Pxydef}.

We next consider the influence of the decoherence between space-time regions on the
dynamics of our system. We begin by discussing the extreme case where~$P(x,y)$ vanishes
identically for~$x$ and~$y$ in different subsystems. Then the action~\eqref{action} splits into the
sum of the actions of the two subsystems, so that the interaction takes place independently in the
two subsystems. In other words, the {\em{subsystems decouple}}. By restricting two different systems
in Minkowski space to~$M_1$ respectively~$M_2$, one can apply the methods of
Section~\ref{seccontinuum} to both subsystems separately. Then each subsystem is described
by an independent continuum limit in terms of a Dirac equation~\eqref{DiracP} coupled to a
classical field~\eqref{Maxwell}. This explains the assumption of an {\em{independent dynamics of
the subsystems}} made in Section~\ref{secindepend}.
We point out that, following the concept that the wave functions generate the causal and
geometric structure of space-time (see the last paragraph in Section~\ref{seccontinuum}),
the decoupling of the subsystems even implies that between the subsystems, the usual causal and
topological structure of Minkowski space ceases to exist.

If we merely assume that~$P(x,y)$ is small for~$x$ and~$y$ in different subsystems,
our action principle~\eqref{action} does describe a coupling of the two subsystems,
which however should be weak.
Keeping in mind that the causal structure of Minkowski space is related to the singularities
of distributions like~\eqref{Psea} on the light cone, and that this singular structure
will be destroyed by decoherence, we know that the coupling of the two subsystems
cannot be described by causal equations formulated in Minkowski space.
Although we have a precise mathematical framework~\eqref{action}, describing the
coupling of the subsystems quantitatively seems very difficult and goes beyond
the scope of this paper. But from the mathematical structure of our action it is already clear that we do not
get contributions from the boundaries of the two subregions. Therefore, instead of
considering the ``layers'' in Figure~\ref{fig1}, it seems more appropriate to draw
each subsystem as many disconnected ``bubbles'' in space-time as shown in Figure~\ref{fig2}.
In view of the continuum limit, each system has an underlying smooth
structure inherited from a corresponding system in Minkowski space, as is indicated in
Figure~\ref{fig2} by the two  ``smooth space-time sheets'' $M^\text{cont}_{1\!/\!2}$.
But of course, this picture should be considered only as a vain attempt to illustrate
an unknown and probably very complicated microscopic structure of space-time
(a maybe more realistic picture will be outlined in Section~\ref{secqurem}).

\subsection{Justification of the Superposition of Fock States} \label{secFsuper}
We now want to verify that expectation values computed with respect to the measurement scalar
product indeed involve superpositions of Fock states. For simplicity, we again consider the situation
for two subsystems~$M_1$ and~$M_2$ and assume that the fermions in each subsystem
are described by an $n$-particle Hartree-Fock state
\[ \psi_1^{(a)} \wedge \cdots \wedge \psi_n^{(a)} \:, \]
whereas the remaining $f-n$ particles describe the Dirac sea. Thus we choose
the one-particle wave functions before microscopic mixing as
$\psi_1^{(a)}, \ldots, \psi_n^{(a)}, \psi_{n+1}^{(a)}, \ldots, \psi_f^{(a)}$,
where the first~$n$ wave functions describe the particles, whereas the other wave functions
form the sea. We again introduce a decoherence between the subsystems by
a unitary transformation of all states in the second subsystem~\eqref{decoher}.

Expectation values~$(\psi_i | \psi_j)$ of the one-particle wave functions with respect to
the measurement scalar product involve a homogenization process, with the result that
wave functions should be identified which differ only by microscopic fluctuations.
More specifically, we should not distinguish between the sea states of the two subsystems. Thus
introducing on~$\H$ the equivalence relation 
\[ \psi \cong \tilde{\psi} \qquad \Longleftrightarrow \qquad ( \psi - \tilde{\psi} \,|\,
\psi - \tilde{\psi} ) = 0\:, \]
we assume that
\[ \psi_j^{(1)} \cong \psi_j^{(2)} \qquad \text{for all~$j=n+1, \ldots, f$}\:. \]
For ease in notation, we make this identification clear simply by omitting the
corresponding superscripts~$^{(1)}$ or~$^{(2)}$.

Under this identification, the many-particle wave function of the whole system becomes
\begin{align}
\Psi &= \psi_1 \wedge \cdots \wedge \psi_f \nonumber \\
&= (\psi_1^{(1)} + \tilde{\psi}_1^{(2)}) \wedge \cdots \wedge (\psi_n^{(1)} + \tilde{\psi}_n^{(2)})
\:\wedge\: (\psi_{n+1} + \tilde{\psi}_{n+1}) \wedge \cdots \wedge (\psi_f + \tilde{\psi}_f)\:.
\label{multiU}
\end{align}
Multiplying out, we obtain many contributions. One of them corresponds to the
many-particle wave function of the first subsystem
\beq \label{mw1}
\psi_1^{(1)} \wedge \cdots \wedge \psi_n^{(1)} \:\wedge\: \psi_{n+1} \wedge \cdots \wedge \psi_f \:,
\eeq
and another is the many-particle wave function of the second subsystem
\begin{align}
\tilde{\psi}_1^{(2)} \wedge &\cdots \wedge \tilde{\psi}_n^{(2)} \:\wedge\: \tilde{\psi}_{n+1} \wedge \cdots \wedge \tilde{\psi}_f = \det U \; \psi_1^{(2)} \wedge \cdots \wedge \psi_n^{(2)} \:\wedge\: \psi_{n+1} \wedge \cdots \wedge \psi_f \nonumber \\
&= \psi_1^{(2)} \wedge \cdots \wedge \psi_n^{(2)} \:\wedge\: \psi_{n+1} \wedge \cdots \wedge 
\psi_f \:. \label{mw2}
\end{align}
All the other contributions involve matrix elements of the unitary operator~$U$.
Similar as explained after~\eqref{Psum}, all these contributions become small if~$U$ is
a random matrix.

We conclude that the measurement process involves the sum of the
many-particle wave functions~\eqref{mw1} and~\eqref{mw2} of the two subsystems.
This justifies Assumption~(B) in Section~\ref{secindepend}.
Moreover, this consideration explains why for measurements one should
work in the Fock space~$\F^\text{eff}_n$ defined by~\eqref{Feffdef}.

\subsection{Describing Fock Superpositions with Random Matrices} \label{secrandom}
In order to make the consideration of the previous section more precise, we now
reformulate its mathematical core in terms of random matrices. Note that~$\SU(f)$
is a compact Lie group, on which we consider a probability measure~$d\mu$.
Then taking the average over a random matrix~$U \in \SU(f)$ corresponds to integrating~$U$
with respect to~$d\mu$. The simplest choice for~$d\mu$ is the normalized
Haar measure~$d\mu_\text{Haar}$ (see for example~\cite[Section~I.5]{broecker+tomdieck}),
but other choices are possible, as will be discussed below.

We first observe that certain products of matrix elements of~$U$ vanish on average.
For simplicity, we consider the Haar measure.
\begin{Lemma} \label{lemmaU}
Suppose that for any~$p$ in the range~$1 \leq p \leq f-1$,
we choose indices~$i_1, \ldots, i_p$ and~$j_1, \ldots, j_p$ with~$i_1 < \cdots < i_p$.
Then
\[ \int_{\SU(f)} U_{i_1 j_1} \cdots U_{i_p j_p}\: d\mu_\text{Haar} = 0 \:. \]
\end{Lemma}
\Proof We let~$k$ be an index which is not contained in $\{i_1, \ldots, i_p\}$
and let~$V$ be the diagonal matrix which has entries one, except that~$V_{i_1 i_1} = V_{k k}=-1$.
Then~$V \in \SU(f)$, and thus a variable transformation shows that the above integral is invariant
under the replacement~$U \rightarrow V U$. But this transformation flips the sign of the
integrand.
\QED \noindent
Applying this lemma to the expression~\eqref{Psum}, we see that the
the fermionic projector~$P(x,y)$ indeed vanishes for~$x$ and~$y$ in different subregions,
if the mean value over~$\SU(f)$ is taken.
The lemma also applies to the contributions obtained by multiplying out~\eqref{multiU}.
It shows that all contributions vanish on average, except for the many-particle wave
functions~\eqref{mw1} and~\eqref{mw2} of the two subsystems.

Since the expectation value of a measurement involves the absolute square of the
wave functions, we also need to integrate the absolute square of the many-particle
wave function~\eqref{multiU} over~$\SU(f)$. We begin with a simple integral involving the
absolute square of one matrix element of~$U$.
\begin{Lemma} \label{lemmascale} For any~$j,k \in \{1,\ldots, f\}$,
\beq \label{av1}
\int_{\SU(f)} |U_{jk}|^2 \: d\mu_\text{Haar} = \frac{1}{f} \:.
\eeq
\end{Lemma}
\Proof By multiplying with suitable unitary operators from the left or the right, we can
arbitrarily change the values of the indices~$j$ and~$k$, without changing the integral.
Thus
\begin{align*}
\int_{\SU(f)} |U_{jk}|^2 \: d\mu_\text{Haar} &= \frac{1}{f^2}\,
\int_{\SU(f)} \sum_{j,k=1}^f |U_{jk}|^2 \: d\mu_\text{Haar} \\
&= \frac{1}{f^2} \,\int_{\SU(f)} \Tr(U^* U) \: d\mu_\text{Haar} = \frac{1}{f}\:,
\end{align*}
concluding the proof.
\QED \noindent
Applying this result to~\eqref{Psum}, we see that decoherence typically scales the kernel of
the fermionic projector down by a factor~$f^{-\frac{1}{2}}$. This quantifies 
that~$P(x,y)$ really becomes small if~$x$ and~$y$ lie in different subregions.

Lemma~\ref{lemmascale} could be generalized to integrals involving the
absolute squares of~$n$ matrix elements, giving the result
\[ \int_{\SU(f)} |U_{j_1 k_1}|^2 \cdots |U_{j_n k_n}|^2 \: d\mu_\text{Haar} \sim \frac{1}{f^n} \qquad
\text{if~$n \ll f$} \:. \]
This shows that every fixed summand obtained by multiplying out~\eqref{multiU}
except for~\eqref{mw1} and~\eqref{mw2} vanishes in the limit~$f \rightarrow \infty$.
Unfortunately, this is not quite good enough, because the number of summands becomes large
if~$f$ increases. Thus in order to estimate the whole sum of terms, we need to use a different
method, which we now explain.

To describe the combinatorics of the wave functions~$\psi^{(1)}_i$ and~$\psi^{(2)}_i$,
we consider a subset~$I^{(1)} \subset \{1, \ldots, n\}$ and take its
complement~$I^{(2)} = \{1, \ldots, n\} \setminus I^{(1)}$. As the case of no particles is trivial,
we may assume that~$n \geq 1$. Suppose that we are interested in the
contribution to~\eqref{multiU} of the form
\[ \Psi \;\asymp\; c(U) \:\Big( \bigwedge\nolimits_{i \in I^{(1)}} \psi^{(1)}_i \Big) \wedge
\Big( \bigwedge\nolimits_{j \in I^{(2)}} \psi^{(2)}_j \Big) \wedge
\psi_{n+1} \wedge \cdots \wedge \psi_f \]
with a complex prefactor~$c(U)$. Multiplying out only the first~$n$ factors in~\eqref{multiU}
and using the definition of~$\tilde{\psi}^{(2)}$ in~\eqref{decoher}, we find that
\[ c(U) = \sign \! \big( I^{(1)} \big)\, \det \!\big( X^{(1)} + U X^{(2)} \big), \]
where~$X^{(a)}$ are the diagonal matrices
\[ (X^{(a)})^i_j = \delta^i_j\: \chi_{I^{(a)} \cup \{n+1, \ldots f\}}(i) \]
(here~$\chi$ is the characteristic function and~$\sign(I)$
defined by
\[ \sign( I ) = (-1)^{i_1 + \cdots + i_g + \frac{g(g+1)}{2}} \]
if we consider~$I$ as the ordered set $I=(i_1, \ldots, i_g)$ with~$1 \leq i_1 < i_2 < \cdots < i_g \leq f$).
Thus to take the average of~$\Psi$ and~$|\Psi|^2$, we need to compute the integrals
\[ \int_{\SU(f)} \det \!\big( X^{(1)} + U X^{(2)} \big)\: d\mu \qquad \text{and} \qquad
\int_{\SU(f)} \Big| \det \!\big( X^{(1)} + U X^{(2)} \big) \Big|^2 d\mu\:, \]
respectively. We get agreement with the Fock space formalism if and only if the following
identities hold:
\begin{align}
\lim_{f \rightarrow \infty} & \int_{\SU(f)} \det \!\big( X^{(1)} + U X^{(2)} \big)\: d\mu =
\delta_{I^{(1)}, \varnothing} + \delta_{I^{(2)}, \varnothing} \label{cr1} \\
\lim_{f \rightarrow \infty} & \int_{\SU(f)} \overline{\det \!\big( X^{(1)} + U X^{(2)} \big)}\:
\det \!\big( \tilde{X}^{(1)} + U \tilde{X}^{(2)} \big) \: d\mu \nonumber \\
&\qquad\qquad\qquad\qquad\qquad\qquad\;\; = c \,\delta_{I^{(1)}, \tilde{I}^{(1)}} \left( \delta_{I^{(1)}, \varnothing} + \delta_{I^{(2)}, \varnothing} \right) .
\label{cr2}
\end{align}
Here the parameter~$c>0$ is an overall constant (to be chosen independent of~$f$), which
can be absorbed into the definition of the scalar product on the Fock space~$\F^\text{eff}_n$.

In the above integrals we worked with a general probability measure~$d\mu$. Let us discuss how it is
to be chosen.
Ultimately, this measure should be determined by analyzing the statistics of the decoherent subsystems
which form according to the action principle~\eqref{action}. Since such an analysis
is not yet available, we must rely on heuristic considerations. One guide line is that~$d\mu$ should
respect all symmetries of the underlying framework. The simplest measure which meets this requirement
is the Haar measure~$d\mu_\text{Haar}$. Writing the above determinant
as~$\det(\1 X^{(1)} + U X^{(2)})$, one sees that the identity matrix is distinguished, and thus
we can choose~$d\mu$ more generally as any measure which is formed of~$U$ and~$\1$.
For example, one could choose~$d\mu$ to be a constant times the measure
\[ |\Tr(\1 + U)|^\alpha \: d\mu_\text{Haar}, \quad
\left|\Tr \!\big( (\1 + U)^\alpha \big) \right| \,d\mu_\text{Haar} \quad \text{or} \quad
|\det(\1 + U)|^\alpha \: d\mu_\text{Haar} \]
with~$\alpha \in \R$, but many other choices are possible (see for example~\cite{mehta}).
The question for which choices of~$d\mu$ the relations~\eqref{cr1} and~\eqref{cr2} hold is an
open problem which will be considered elsewhere.

We finally remark that the above considerations immediately generalize to more than two
subsystems, if one replaces the term~$\det(X^{(1)} + U X^{(2)})$ by
\[ \det \!\big( X^{(1)} + U_2 X^{(2)} + \cdots + U_L X^{(L)} \big) \]
and integrates over all random matrices~$U_2, \ldots, U_L \in \SU(f)$.

\section{Second Quantization of the Bosonic Field} \label{secbosonic}
In Section~\ref{secdecind} we considered two decoherent subsystems~$M_1$ and~$M_2$
and saw that by analyzing each subsystem in the continuum limit, we could describe the
dynamics by the Dirac equation coupled to a classical field.
Taking a finite number of such decoherent subsystems, the whole dynamics is described by
several classical fields, one for each subsystem. In this chapter we will show that the
resulting framework indeed allows for the description of second quantized bosonic fields.

For simplicity, we merely consider an electromagnetic field (the generalization
to other bosonic fields is straightforward). We subdivide Minkowski space into~$L$ disjoint
regions~$M_1, \ldots, M_L$, which are again assumed to be fine-grained.
Similar to~\eqref{Pxydef} and~\eqref{Psplit}, the fermionic projector can be written as
\beq \label{Psub}
P(x,y) = -\sum_{j=1}^f |\psi_j(x) \Sr \Sl \psi_j(y) | \qquad \text{with} \qquad
\psi_j = \sum_{a=1}^L \psi_j^{(a)} , \;\; \psi_j^{(a)} := \psi_j \: \chi_{M_a} \:,
\eeq
where~$f$ is a large number which tends to infinity if the ultraviolet regularization
is removed. As in Section~\ref{secdecind}, we arrange by unitary
transformations of the form~\eqref{decoher} that the subsystems are decoherent.
Considering each subsystem in the continuum limit, we obtain
similar to~\eqref{DiracP} and~\eqref{Maxwell} the Dirac-Maxwell equations
\beq \begin{split} \label{DMsub}
\Big( i \gamma^j (&\partial_j - ie A^{(a)}_j) - m \Big) P^{(a)}(x,y) = 0 \\
\partial_j^{\;\,k} A^{(a)}_k - \Box A^{(a)}_j &= e \sum_{k=1}^{n_f} \Sl \psi^{(a)}_k | \gamma_j \psi^{(a)}_k \Sr
-e \sum_{l=1}^{n_a} \Sl \phi^{(a)}_l | \gamma_j \phi^{(a)}_l \Sr\:.
\end{split}
\eeq
We note for clarity that according to~\eqref{Psplit}, the wave functions~$\psi^{(a)}_k$
and~$\phi^{(a)}_l$ are obtained by restriction to a subregion $M_a \subset M$
of space-time. But it is reasonable to assume that these wave functions are macroscopic
in the sense that they can be extended smoothly to the whole Minkowski space. Similarly, we
assume that the potentials~$A^{(a)}$ and the fermionic projectors~$P^{(a)}$ are defined
on a whole sheet~$M^\text{cont}_a$ of Minkowski space (see Figure~\ref{fig2}).

\subsection{Describing a Second Quantized Free Bosonic Field} \label{secquEM}
In order to get a simple connection to standard textbooks like~\cite{zeidler1, landau4}, we begin
with a free electromagnetic field (i.e.\ the situation where no fermionic particles or anti-particles are
present). Furthermore, to avoid the technical issues involved in taking an infinite volume limit,
we restrict attention to the situation in finite spatial volume by considering a
three-dimensional box of length~$\ell$ with periodic boundary conditions.
Working in the Coulomb gauge~$\text{div} \vec{A}=0$, the Maxwell equations reduce to
the ordinary wave equation for each component of the vector potential,
\[ \Box \vec{A}(t, \vec{x}) = 0\:, \]
whereas the electric potential~$A^0$ can be arranged to vanish identically.
Decomposing~$\vec{A}$ into Fourier modes of momentum~$\vec{k} \in (2 \pi \Z /\ell)^3$,
\[ \vec{A}(t,\vec{x}) = \sum_{\vec{k}} \left( \vec{a}(t, \vec{k}) \,e^{i \vec{k} \vec{x}} 
+  \overline{\vec{a}(t, \vec{k})} \,e^{-i \vec{k} \vec{x}} \right) \:, \]
the Maxwell equations are solved by
\[ \vec{a}(t, \vec{k}) = \vec{a}(\vec{k}) \: e^{-i \omega t} \quad \text{with} \quad
\omega := |\vec{k}| \:, \]
whereas the Coulomb gauge gives rise to the transversality condition~$\vec{k} \cdot
\vec{a}(\vec{k}) = 0$ (see~\cite[Chapter~I, \S2]{landau4}). The two linearly independent solutions of this transversality condition
correspond to the two polarizations of the electromagnetic wave; we denote them
by an index~$\beta=1,2$. Introducing the canonical field variables
\[ q_\beta(\vec{k}) = \frac{1}{4 \pi} \left( a_\beta(\vec{k}) + \overline{a_\beta(\vec{k})} \right) \:,
\qquad p(\vec{k}) = \frac{d}{dt} q_\beta(\vec{k}) = -\frac{i \omega}{4 \pi}
\left( a_\beta(\vec{k}) - \overline{a_\beta(\vec{k})} \right) , \]
the energy~$H$ of the classical electromagnetic field becomes (for details
see~\cite[Chapter~I, \S2]{landau4})
\beq \label{Hclassical}
H = \sum_{\vec{k} \in (2 \pi \Z /\ell)^3} \;\sum_{\beta=1,2}
\frac{1}{2} \left( p_\beta(\vec{k})^2 + \omega^2 q_\beta(\vec{k})^2 \right) .
\eeq
Here each summand is the Hamiltonian of a harmonic oscillator. Thus we have rewritten the
classical electromagnetic field as an infinite collection of classical harmonic oscillators.

The second quantization of the electromagnetic field corresponds to
quantizing each harmonic oscillator as in standard quantum mechanics
(see for example~\cite[Part~I, Section~1.2]{zeidler1}). We proceed by discussing the
connection between the classical and the quantum dynamics in detail, for
simplicity for a single harmonic oscillator of frequency~$\omega$.
Thus our starting point is the classical Hamiltonian
\beq \label{hfosci}
h(p,q) = \frac{1}{2} \left( p^2 + \omega^2 q^2 \right) .
\eeq
Here~$q$ and~$p$ are the canonical variables, which together form the
classical phase space~${\mathcal{P}} = \{ (p,q) \text{ with } p,q \in \R\}$.
The classical dynamics is described by Hamilton's equations
\[ \frac{d}{dt} q = \frac{\partial h}{\partial p} = p \:,\qquad
\frac{d}{dt} p = -\frac{\partial h}{\partial q} = -\omega^2 q\:. \]
A solution~$(p(t), q(t))$ describes a classical trajectory.
Solving Hamilton's equations, the classical dynamics describes a rotation in phase space,
\beq \label{flow}
\begin{pmatrix} p(t) \\ \omega q(t) \end{pmatrix} 
= \begin{pmatrix} \cos \omega t & -\sin \omega t \\ \sin \omega t & \cos \omega t \end{pmatrix}
\begin{pmatrix} p(0) \\ \omega q(0) \end{pmatrix} .
\eeq
In order to get a setting similar to that in quantum theory, we next consider on
phase space complex-valued functions~$\psi(p,q)$, referred to as ``classical wave functions.''
Introducing the scalar product
\beq \label{spclass}
\la \psi | \phi \ra_\cl = \iint_{\R \times \R} \overline{\psi(p,q)} \phi(p,q)\: dp\, dq\:,
\eeq
the classical wave functions form a Hilbert space~$({\mathcal{H}}_\cl, \la .|. \ra_\cl)$.
The phase flow~\eqref{flow} induces a flow on~${\mathcal{H}}$, which is described most
conveniently by the {\em{time evolution operator}}~$U_\text{class}$ defined by
\beq \label{Uclass}
\Big(U_\cl(t) \,\psi \Big)(p(t),q(t)) = \Big(U_\cl(0) \,\psi \Big)(p(0),q(0)) \:.
\eeq
It is a unitary operator on~${\mathcal{H}}_\cl$.
Before going on, we remark that in classical physics one usually works instead of
complex functions with positive functions or densities on phase space.
Working with complex-valued functions and the scalar product~\eqref{spclass}
seems unusual but will be very useful for the following considerations.
In a somewhat different context, the Hilbert space~$({\mathcal{H}}_\cl, \la .|. \ra_\cl)$
is also used in geometric quantization for the the so-called prequantization
(see~\cite[Section~5.2]{woodhouse}).

For the quantization of the oscillator, one replaces the canonical variables~$p$ and~$q$
by self-adjoint operators~$P$ and~$Q$ which act on a complex Hilbert space~$({\mathcal{H}}, \la .|. \ra)$
and satisfy the canonical commutation relations~$[P, Q]=-i$. The physical system is now
characterized by a state~$\Psi \in {\mathcal{H}}$. The dynamics is described by the Schr\"odinger
equation
\beq \label{schroding}
i \partial_t \Psi = H \Psi \qquad \text{with} \qquad H = \frac{1}{2} \left( P^2 + \omega^2 Q^2 \right) .
\eeq
It is most common to represent~${\mathcal{H}}$ as the space of square integrable functions
with the inner product
\beq \label{spqu}
\la \Psi | \Phi \ra = \int_\R \overline{\Psi(q)} \Phi(q)\:dq\:,
\eeq
and to choose the operators~$Q$ and~$P$ as
\[ (Q \psi)(q) = q\, \psi(q) \qquad \text{and} \qquad P = -i \frac{d}{dq}\:. \]
Integrating the Schr\"odinger equation gives rise to the unitary time evolution operator
\beq \label{Uqu}
U(t) = e^{-i t H}\::\: \H \rightarrow \H \::\: \Psi(0, q) \mapsto \Psi(t, q)\:.
\eeq

With the above formulation we expressed both the classical and the quantum
dynamics by a unitary time evolution operator acting on a Hilbert space
(see~\eqref{Uclass}, \eqref{spclass} and~\eqref{Uqu}, \eqref{spqu}). But the
time evolution operators have a completely different form. Furthermore, the
Hilbert spaces are different, because the ``classical wave functions'' depend on both~$q$ and~$p$.
In the quantized theory, however, the Heisenberg uncertainty principle prevents~$P$
and~$Q$ from being simultaneously measurable, as is reflected mathematically by the fact
that they correspond to non-commuting operators. Since in the classical theory,
position and momentum can be chosen independently, there is much more freedom
to choose the initial wave function~$\psi(p,q)$ than in quantum theory, where choosing~$\Psi(q)$
automatically determines the corresponding wave function in momentum space.
This raises the question if for a given quantum wave function~$\Psi(q)$ we can choose a
corresponding classical wave function~$\psi(p,q)$ such that the classical dynamics
of~$\psi$ as described by~\eqref{Uclass} coincides with the time evolution
of the quantum wave function~\eqref{Uqu}.
While the general answer to this question is no, it turns out that for the harmonic
oscillator this correspondence can indeed be made:

\begin{Lemma} {\bf{(Correspondence between classical and quantum dynamics)}} \label{lemmacorr}
Consider the classical harmonic oscillator~\eqref{hfosci} with dynamics~\eqref{flow}, \eqref{Uclass}
and the corresponding quantum harmonic oscillator with the dynamics described
by the Schr\"odinger equation~\eqref{schroding} and~\eqref{Uqu}.
Then there is an isometric embedding~$\iota \::\: {\mathcal{H}} \rightarrow {\mathcal{H}}_\cl$
which maps the quantum evolution to a corresponding classical evolution, in the sense that
\beq \label{Ucorr}
U_\cl(t) \:\iota = \iota \:U(t) \:\: e^{i \omega t/2} \qquad \text{for all~$t \in \R$}\:.
\eeq
Moreover, there are differential operators~$H_\cl$, $P_\cl$ and~$Q_\cl$ in~${\mathcal{H}}_\cl$
such that
\[ H_\cl  \:\iota = \iota \: H \:,\qquad P_\cl  \:\iota = \iota \: P \qquad \text{and}
\qquad Q_\cl  \:\iota = \iota \: Q \:. \]
\end{Lemma} \noindent
We point out that the factor~$e^{i \omega t/2}$ in~\eqref{Ucorr} corresponds to the zero point
energy of the quantum harmonic oscillator.
This factor modifies the wave functions only by a joint global phase,
without an influence on any observations or expectation values.
\Proof[Proof of Lemma~\ref{lemmacorr}]
We choose an orthonormal eigenvector basis~$\Psi_n$ of the Hamiltonian in~\eqref{schroding}
 (see for example~\cite[Section~3.1]{schwabl1})
 \[ H \Psi_n = \left(n + \frac{1}{2} \right) \omega \Psi_n \:,\qquad n=0,1,\ldots . \]
 Writing the Hamiltonian as~$H=\omega (a^\dagger a+\frac{1}{2})$ with the
annihilation and creation operators
\beq \label{crean}
a = \frac{1}{\sqrt{2 \omega}} \left( \omega q + \frac{d}{dq} \right) ,\qquad
a^\dagger = \frac{1}{\sqrt{2 \omega}} \left( \omega q - \frac{d}{dq} \right) ,
\eeq
the eigenvectors can be obtained by acting with the creation operators on the
ground state,
\beq \label{Psidef}
\Psi_0 = c_0\: \exp \left( -\frac{\omega q^2}{2} \right) \qquad \text{and} \qquad
\Psi_n = c_n \:(a^\dagger)^n \,\Psi_0 \:,
\eeq
were the~$c_n$ are positive normalization constants. From~\eqref{Uqu} it follows immediately that
\beq \label{Uquc}
U(t)\, e^{i \omega t/2}\, \Psi_n = e^{-i n \omega t }\: \Psi_n \:.
\eeq

In order to define the mapping~$\iota$, it suffices to associate to every eigenfunction~$\Psi_n$
a corresponding classical wave functions~$\psi_n \in {\mathcal{H}}_\cl$
(then~$\iota$ is determined uniquely by linearity and continuity).
First, in order to write the classical dynamics in a
simpler form, we rescale the momentum variable by introducing the new phase space variables
\[ x = q \qquad \text{and} \qquad y = \frac{p}{\omega}\:. \]
Setting~$z=x+iy$, the time evolution operator~\eqref{Uclass} becomes
\beq \label{Ucl}
(U_\cl(t) \psi)(z) = \psi(e^{i \omega t} z) \:.
\eeq
We now define the ``classical annihilation and creation operators'' on~${\mathcal{H}}_\cl$ by
\[ a_\cl = \frac{1}{2} \left( a_x + i a_y \right) ,\qquad
a^\dagger_\cl = \frac{1}{2} \left( a^\dagger_x + i a^\dagger_y \right) , \]
where~$a_x$ and~$a_x^\dagger$ are given in analogy to~\eqref{crean} by
\[ a_x = \frac{1}{\sqrt{2 \omega}} \left( \omega x + \frac{d}{dx} \right) ,\qquad
a_x^\dagger = \frac{1}{\sqrt{2 \omega}} \left( \omega x - \frac{d}{dx} \right) , \]
whereas the subscript~$y$ refers similarly to the variable~$y$.
We introduce the wave functions~$\psi_n$ by
\beq \label{psidef}
\psi_0 = c_0^2 \: \exp \left( -\frac{\omega (x^2+y^2)}{2} \right) \qquad \text{and} \qquad
\psi_n = c_n\:(a^\dagger_\cl)^n \,\psi_0 \:.
\eeq

Let us verify that the resulting mapping~$\iota$ has the required properties. First,
it is obvious from their definition~\eqref{psidef} that the functions~$\psi_n$ are
orthonormal in~${\mathcal{H}}_\cl$, and thus~$\iota$ is indeed an isometric embedding.
Using a polar decomposition~$z = r e^{i \varphi}$, a short calculation shows that
\[ \left[ i \partial_\varphi, a^\dagger_\cl \right] = a^\dagger_\cl \:. \]
Applying this relation in~\eqref{psidef} and using that~$\psi_0$ is radially symmetric, we obtain
\[ i \partial_\varphi \psi_n = n \psi_n \qquad \text{and thus} \qquad
\psi_n(z) = e^{-i n \varphi} \: \phi_n(r) \]
with radially symmetric functions~$\phi_n$. Thus the classical dynamics~\eqref{Ucl}
implies that
\[ U_\cl(t)\: \psi_n = e^{-i \omega n t}\: \psi_n \:. \]
Comparing with~\eqref{Uquc} proves~\eqref{Ucorr}.

In order to construct the operators~$H_\cl$, $P_\cl$ and~$Q_\cl$, we first note that
both the classical and quantum annihilation and creation operators satisfy the
canonical commutation relations
\[ \left[ a^\dagger_\cl, a_\cl \right] = 1 \qquad \text{and} \qquad
\left[ a^\dagger, a \right] = 1 \:, \]
and in view of~\eqref{Psidef} and~\eqref{psidef} they correspond to each other
in the sense that
\[ a_\cl \,\iota = \iota \,a \qquad \text{and} \qquad a^\dagger_\cl \,\iota = \iota\, a^\dagger \:. \]
Thus expressing the operators in~${\mathcal{H}}$ in terms of~$a$ and~$a^\dagger$, we obtain
the corresponding ``classical'' operators simply by adding subscripts. Thus we set
\[ H_\cl = \omega \left(a^\dagger_\cl a_\cl+\frac{1}{2} \right) \]
and
\[  Q_\cl = \frac{1}{\sqrt{2 \omega}} \left( a_\cl + a_\cl^\dagger \right) \:,\qquad
P_\cl = -i \,\sqrt{\frac{\omega}{2}} \left( a_\cl - a_\cl^\dagger \right) . \]
This concludes the proof.
\QED
We remark that the mapping~$\iota$ appears in the mathematical physics literature
as the so-called Bargmann transform (see~\cite[Section~4.3]{weaver}). But to our knowledge,
it has not been used to get a connection between the classical and quantum dynamics.

The above lemma shows that by choosing the ``classical wave function'' $\phi \in {\mathcal{H}}_\cl$
appropriately, we can arrange that the classical dynamics reproduces any quantum dynamics.
In simpler terms, the quantum dynamics of the harmonic oscillator can be recovered as
a special case of the classical dynamics. However, for making this correspondence, we had
to take a somewhat unusual point of view and work on classical phase space with complex-valued
functions and the scalar product~\eqref{spclass}.
To us, Lemma~\ref{lemmacorr} is useful because it makes it possible to {\em{approximate
a quantum state by a finite number of classical trajectories}}, if to every classical trajectory
we associate a complex number. This can be seen as follows: Suppose that we want to
describe a quantum state~$\Psi \in \H$. According to Lemma~\ref{lemmacorr}, this state
has the same dynamics as the classical wave function~$\psi := \iota \Psi \in \H_\text{class}$.
For any~$L \in \N$ and an index~$a =1, \ldots, L$, we now choose points~$(p^{(a)}, q^{(a)})$
in phase space together with complex coefficients~$\phi(a)$ which approximate~$\psi$ in the
sense that
\beq \label{discapprox}
\sum_{a=1}^L \phi(a)\: \delta \big(p - p^{(a)} \big)\: \delta \big( q-q^{(a)} \big) 
\xrightarrow{L \rightarrow \infty} \psi(p,q) \:,
\eeq
with convergence in the distributional sense. For these discrete configurations, we can make sense
of the scalar product~\eqref{spclass} by setting
\[ \langle (p^{(a)}, q^{(a)}, \phi(a)) | (\tilde{p}^{(b)}, \tilde{q}^{(b)}, \tilde{\phi}(b))  \rangle
= \sum_{a,b} \delta_{p^{(a)}, \tilde{p}^{(b)}}\: \delta_{q^{(a)}, \tilde{q}^{(b)}}\;
\overline{\phi(a)} \tilde{\phi}(b)\:. \]
Thus by choosing~$L$ sufficiently large,
we can approximate the quantum dynamics of~$\Psi$ by a complex valued function~$\phi$
defined on a finite number of classical trajectories.

Before going on, we point out that the scalar product~\eqref{spclass} is invariant under the
local phase transformations
\beq \label{lgtphase}
\phi(p,q) \rightarrow e^{i \varphi(p,q)}\: \phi(p,q)\:,
\eeq
where~$\varphi$ is a real-valued function on phase space. Thus the phase of the functions
in~$\H_\text{class}$ has no physical relevance. What counts is only
the {\em{relative phase}} when taking superpositions
of two wave functions~$\psi, \phi \in {\mathcal{H}}_\cl$. Similarly, in the discrete
approximations in~\eqref{discapprox}, the phase of the function~$\phi(a)$
can be changed by
\[ \phi(a) \rightarrow e^{i \varphi(a)} \phi(a) \quad
\text{under the constraint} \quad (p^{(a)}, q^{(a)}) = (p^{(b)}, q^{(b)}) 
\Longrightarrow \varphi(a)=\varphi(b)\:. \]

In the remainder of this section, we extend the above considerations on the harmonic
oscillator to the Hamiltonian of the electromagnetic field~\eqref{Hclassical}.
By taking tensor products, the result of Lemma~\ref{lemmacorr} immediately extends to 
a collection of harmonic oscillators as in~\eqref{Hclassical}.
It then states that by considering suitable complex-valued functions on the
set of all classical field configurations, one can reproduce the full dynamics of the free
second-quantized field. Using an approximation argument similar to~\eqref{discapprox},
it suffices to consider a finite number of classical field configurations.
Thus the remaining task is to associate to every classical field configuration a complex number~$\phi(a)$.
Let us return to the setting of decoherent subsystems in the continuum limit~\eqref{DMsub}.
Then every subsystem involves a classical electromagnetic potential~$A^{(a)}$.
In the considered case without fermions, the field equations reduce to the free Maxwell equations, i.e.\
again in the Coulomb gauge
\beq \label{Maxsub}
\Box \vec{A}^{(a)}(t, \vec{x}) = 0
\eeq
and~$(A^{(a)})^0(t, \vec{x}) = 0$. Moreover, we have the Dirac equation for the fermionic
projector, which according to~\eqref{particles} consists only of the sea states,
\beq \label{Dirsub}
\left( i \gamma^j (\partial_j - ie A^{(a)}_j) - m \right) P^\text{sea}(x,y) = 0 \qquad \text{if~$x \in M_a$}\:.
\eeq
Having an ultraviolet regularization in mind, the number~$f$ of sea states is finite,
so that~$P^\text{sea}$ can be written in the form~\eqref{Psub}.
For a given solution~$\check{A}$ of the free Maxwell equations, we now introduce the following
{\em{reference system}}. The causal perturbation expansion distinguishes
a subspace of the solution space of the Dirac equation as being formed by the sea states.
Selecting the $f$-dimensional subspace of the sea states which is compatible with our regularization
and choosing an orthonormal basis~$\check{\psi}_1, \ldots, \check{\psi}_f$ of this subspace, we can
introduce the many-particle wave function~$\check{\psi}$ of our reference system by
\[ \check{\Psi} = \check{\psi}_1 \wedge \cdots \wedge \check{\psi}_f \:. \]
As in~\eqref{psiphase}, the freedom in choosing the orthonormal basis implies that~$\check{\Psi}$
is determined only up to a phase.
Now suppose that~$\check{A}$ coincides with the electromagnetic potential~$A_a$ in one of our subsystems.
Then the wave functions~$\psi^{(a)}_1, \ldots, \psi^{(a)}_f$ obtained by restricting the
wave functions of the fermionic projector of the whole system to the subsystem~$M_a$
span the same subspace of the Dirac solutions as the vectors~$\check{\psi}_1, \ldots, \check{\psi}_f$
(probably after suitably modifying the solutions on the microscopic scale or modifying
the regularization; a technical issue which for simplicity we will ignore here).
Hence the corresponding many-particle wave function~\eqref{Psisub} coincides
with~$\check{\Psi}$ up to a complex number,
\beq \label{Psia}
\Psi^{(a)} = \phi(a)\: \check{\Psi} \qquad \text{with} \qquad \phi(a) \in \C\:.
\eeq
In this way, we have associated to the field configuration~$A_a$ a complex number~$\phi(a)$.

Let us consider the phase freedom. 
As noted above, the phase of the wave function~$\check{\Psi}$ depends on the
choice of the basis~$\check{\psi}_1, \ldots, \check{\psi}_f$.
Similarly, by transforming the orthonormal basis~$\psi_1, \ldots, \psi_f$ of the image of~$P$,
we can also change the phase of~$\Psi^{(a)}$ arbitrarily. Thus~\eqref{Psia} is well-defined only
up to a phase. Now suppose that~$\check{A}$ also coincides with the electromagnetic potential~$A_b$
of another subsystem. Then writing the many-particle wave function of the new subsystem
as~$\Psi^{(b)} = \phi(b)\: \check{\Psi}$,
transforming the bases~$\check{\psi}_1, \ldots, \check{\psi}_f$ or~$\psi_1, \ldots, \psi_f$
changes the phase of both~$\phi(a)$ and~$\phi(b)$ in the same way. Thus
the relative phase of~$\phi(a)$ and~$\phi(b)$ is well-defined. In other words, the
complex-valued function~$\phi$ is uniquely defined up to the transformations
\beq \label{lgtfield}
\phi(a) \rightarrow e^{i \varphi(a)} \phi(a) \qquad
\text{under the constraint} \qquad A_a = A_b \Longrightarrow \varphi(a)=\varphi(b)\:.
\eeq
These transformations can be regarded as local phase transformations on the classical field
configurations, just as explained after~\eqref{lgtphase} for one harmonic
oscillator on phase space.

We conclude that the above construction indeed yields a complex-valued wave function~$\phi(a)$,
$a=1,\ldots, L$, defined on the classical field configurations~$\{ A^{(1)}, \ldots, A^{(L)} \}$
of the subsystems. It is uniquely determined up to the local phase transformations~\eqref{lgtfield}.
These results make it possible to {\em{approximate a general state of the bosonic Fock space
by our decoherent subsystems}}, as the following consideration shows:
According to Lemma~\ref{lemmacorr}, the dynamics of a given bosonic Fock state can
be described by a complex-valued wave function~$\phi$ on the classical field configurations.
By considering similar to~\eqref{discapprox}
a sequence of systems where the number of decoherent subsystems tends to infinity,
we can approximate~$\phi$ by a finite collection of classical field configurations
$\{ A^{(1)}, \ldots, A^{(L)} \}$ and a corresponding complex-valued functions~$\phi(a)$.
By suitably adjusting
the phases of the sea states~$\psi^{(a)}_1, \ldots, \psi^{(a)}_f$ (for given reference
systems~$\check{\Psi}$), we can arrange that the function~$\phi(a)$ satisfies~\eqref{Psia}.
Then the wave functions~$\psi_1, \ldots, \psi_f$ of the whole system encode the
classical potentials $\{ A^{(1)}, \ldots, A^{(L)} \}$ as well as the complex-valued function~$\phi(a)$,
which together approximate the given bosonic Fock state.

\subsection{Describing a Second Quantized Fermion-Boson System} \label{secqubos}
We now combine the considerations of the previous section 
with the constructions of Chapter~\ref{secmix} to obtain
a unified framework for describing second quantized fermions and bosons.
We again consider $L$ decoherent subsystems in the continuum limit~\eqref{DMsub}.
According to~\eqref{particles}, we can split up the fermionic projector into the
particle- and anti-particle- as well as the sea states. We begin for clarity in the situation
without pair creation where the numbers~$n_a$ and~$n_f$
of particles and anti-particles are constant and coincide in all subsystems (this constraint
will be removed below). Then setting~$n=n_a-n_f$, the corresponding
many-particle wave functions of the subsystems~\eqref{Psisub} can be decomposed as
\beq \label{Psifermbos}
\Psi^{(a)} = \Big( \psi_1^{(a)} \wedge \cdots \wedge \psi_{n_f}^{(a)}
\wedge \phi_1^{(a)} \wedge \cdots \wedge \phi_{n_a}^{(a)} \Big)
\wedge \Big[ \psi_{n+1} \wedge \cdots \wedge \psi_f \Big] \:.
\eeq
Here the round brackets can be regarded as the fermionic wave function of the particles and anti-particles.
As explained in Section~\ref{secFsuper}, measurements involve superpositions of these
many-particle wave functions, so that it is reasonable to regard the round brackets in~\eqref{Psifermbos}
as a vector in the fermionic Fock space~$\F^\text{eff}$.
Likewise, the square brackets in~\eqref{Psifermbos} describe the sea.
The construction~\eqref{Psia} yields a corresponding complex wave function~$\phi(a)$
on the classical field configurations, which can be used to describe the dynamics of
a second quantized bosonic field. In this way, we have extracted from the fermionic projector
both a fermionic and a bosonic quantum field.

We now give a construction which avoids the splitting of the
many-particle wave function into the particle/anti-particle component and the sea component.
Apart from being simpler and cleaner, this construction has the advantage of working just as well
for fully interacting systems, including pair creation or annihilation processes.
Recall that in~\eqref{Psia} we compared the many-particle wave function~$\Psi^{(a)}$ of
our subsystem with the wave function~$\check{\Psi}$ of a ``reference system'' having the same
classical field configuration. The proportionality factor~$\phi(a)$ then gave us the
desired complex-valued function~$\phi$ on the classical field configurations.
Giving up the requirement that the vector space~${\mathcal{H}}_\cl$ should be
represented by complex-valued functions, one can work instead of~$\phi(a)$
with the corresponding vector~$\Psi^{(a)} \in \F^\text{eff}_f$.
This has no effect on superpositions, because the complex coefficients~$\phi(a)$ and~$\phi(b)$
can be linearly combined only if the corresponding
classical field configurations~$A^{(a)}$ and~$A^{(b)}$ coincide.
But then the corresponding Fock vectors~$\Psi^{(a)}, \Psi^{(b)} \in \F^\text{eff}_f$ are linearly
dependent, so that taking their linear combination has the same effect as taking the linear
combinations of the coefficients~$\phi(a)$ and~$\phi(b)$.
This leads us to replace the complex-valued function~$\phi(a)$
constructed in~\eqref{Psia} by a mapping with values in~$\F^\text{eff}_f$,
\beq \label{phidef}
\phi \::\: \{1, \ldots, L \} \rightarrow \F^\text{eff}_f \::\:
a \mapsto \psi^{(a)}_1 \wedge \cdots \wedge \psi^{(a)}_f\:.
\eeq
In the setting involving particles and anti-particles~\eqref{Psifermbos}, this mapping has the nice
property that it involves at the same time the fermionic wave functions of the particles and anti-particles.
In free field theory, it can be thought of as the tensor product of
a fermionic and a bosonic Fock state. As desired, two such tensor states are linearly
dependent only if both the fermionic and bosonic parts are.
Superpositions of these tensor states can be justified exactly as explained in Section~\ref{secFsuper}.
In a fully interacting system, the mapping~\eqref{phidef} can no longer be decomposed
into a fermionic and a bosonic part, in agreement with the fact that in interacting quantum
systems the bosonic and fermionic Fock spaces are coupled together and ``mixed'' by the Hamiltonian.
Even in this in general very complicated situation, the mapping~$\phi$ gives a conceptually simple
description of the whole system.

\subsection{Remarks and Outlook} \label{secqurem}
To avoid confusion, we point out that the constructions in this chapter are not equivalent
to the canonical quantization of the bosonic field. In particular, we
do not get the physical equations for second quantized fields. Instead, we merely show
that the dynamics of {\em{free second quantized bosonic fields}} can be mimicked
by an ensemble of decoherent subsystems, each with a classical dynamics.
However, we do not get a justification nor an explanation for the
physical assumption that electromagnetic wave modes should behave like quantum mechanical
oscillators. But we show that this assumption is not in conflict with the framework of
the fermionic projector. In particular, it is possible to describe {\em{entangled bosonic states}}.

In order to explain why we do not even attempt to reproduce the physical equations
for second quantized fields, we now briefly outline how interacting quantum field theory
should be formulated in the framework of the fermionic projector.
Recently, this formulation of quantum field theory has been worked out in detail
for a system involving an axial field~\cite{sector}. The general strategy is as follows.
Instead of quantizing the classical field equations, we describe the interaction and the
dynamics of the system by the action principle~\eqref{action}.
In the continuum limit, the Euler-Lagrange equations corresponding to this action
principle give rise to the Dirac equation coupled to classical bosonic field equations
(cf.~\eqref{DiracP} and~\eqref{Maxwell}).
Treating this coupled system of nonlinear partial differential equations in a perturbation
expansion gives rise to all the Feynman diagrams of perturbative quantum field
theory (see~\cite[Section~8.4]{sector}). In particular, this gives agreement with
the high-precision tests of quantum field theory.
We remark that we get additional small corrections to the field equations which are absent in
perturbative quantum field theory; the interested reader is referred to~\cite[Section~8.2
and~8.3]{sector}.

Since the quantitative aspects are respected, it remains to explain the
particular effects of quantized fields. The present paper is concerned with entanglement
and shows that entangled fermionic and bosonic states can be described in the framework
of the fermionic projector. For other quantum effects related to the measurement problem and
the wave-particle duality see also~\cite[Section~4]{rev}. Putting these results together, it seems to us that
the framework of the fermionic projector is in agreement with all effects
and predictions of quantum field theory (except for the additional corrections discussed
in~\cite[Sections~8.2 and~8.3]{sector}). But this agreement cannot be stated in terms
of a mathematical equivalence, partly because standard quantum field theory at present has no
fully convincing mathematical formulation. Also, many difficulties of quantum field theory
clearly remain unsettled in our approach. Thus many conceptual and technical issues need to
be debated in the future.

We finally point out that the agreement with free quantized fields in Section~\ref{secquEM}
is obtained only in the limit when the number of subsystems tends to infinity.
Thus even for describing the quantum oscillations of the harmonic oscillator corresponding
to one mode of the electromagnetic field, one needs to consider a large number of
decoherent subsystems. Although this seems possible in principle, it seems hard to
imagine that decoherence should really lead to a ``fragmentation'' of space-time into many
disjoint regions with an independent dynamics. This raises the question
whether the microscopic mixing of decoherent subsystems might not be a too simple picture for
understanding the mechanisms of space-time on a small scale. Indeed, it might be more
appropriate to replace this picture by a more general concept which we now explain in words.
Recall that in Section~\ref{secdecind} the decoherence of subsystems was introduced by
inserting a unitary transformation into the fermionic projector,
\beq \label{nodecoher}
P(x,y) = -\sum_{j,k=1}^f U_{jk} \:| \psi^{(2)}_j(x) \Sr \Sl \psi^{(1)}_k(y) | \:,
\eeq
if~$x$ and~$y$ are in different subsystems (see~\eqref{Psum}).
Since~$P(x,y)$ is a $4 \times 4$-matrix, a dimensional argument shows that there
is a large class of operators~$U \in \SU(f)$ which do not affect the form of~$P(x,y)$.
In order to make use of this additional freedom, we replace~$U$ by a family of local unitary
transformations~$U(x) \in \SU(f)$, which brings the fermionic projector to the more general form
\beq \label{PUU}
P(x,y) = -\sum_{j,k,l=1}^f U_{jk}(x) \:U^{-1}_{kl}(y) \:| \psi_j(x) \Sr \Sl \psi_l(y) | \:.
\eeq
By dividing~$M$ into subregions and choosing~$U(x)$ to be constant on each subregion,
we get back to the setting of Section~\ref{secdecind}. But if~$U(x)$ is not a piecewise
constant function, the situation is more involved. Namely, for any space-time points~$x$ and~$y$,
it is possible that the transformation~$U_{jk}(x) \:U^{-1}_{kl}(y)$ has no effect on~$P(x,y)$
(similar as discussed in~\eqref{nodecoher}); in this case the pair~$(x,y)$ is said to be
coherent. Another possibility is that the transformation~$U_{jk}(x) \:U^{-1}_{kl}(y)$
leads to cancellations in the sum so that~$P(x,y)$ is very small (similar as explained after~\eqref{Psum}),
in which case the pair~$(x,y)$ is said to be decoherent.
This notion of decoherence again gives a relation between space-time points.
But in contrast to the situation in Section~\ref{secdecind}, this relation is no longer transitive;
for example, it is possible that the pairs~$(x,y)$ and~$(y,z)$ are coherent, but the
pair~$(x,z)$ is decoherent. As a consequence, decoherence no longer gives rise to
a decomposition of space-time into subregions. But for any fixed space-time point~$x$,
one can form the set~$M(x)$ of all space-time points which are coherent to~$x$.
This set can then be divided into subsets~$M_j(x)$ by the condition that any
two points~$y,z \in M_j(x)$ should be coherent to each other.
On the sets~$M_j(x)$, one can then again consider the continuum limit
to obtain for example the Dirac-Maxwell system~\eqref{DiracP}, \eqref{Maxwell}.
Thus on the coherent space-time points one again gets a description involving classical field equations.
Decoherent pairs of space-time points, on the other hand, are not connected by our action principle.
We remark that it is also conceivable that two space-time points are partially decoherent in the sense that
there are cancellations in the sums~\eqref{PUU}, but without~$P(x,y)$ being very small.
We expect that such a partial decoherence would yield a large contribution to the action
and should thus be avoided by our action principle. 
Then the resulting structure resembles the situation in Section~\ref{secdecind} in that
we obtain decoherent subsystems with an independent dynamics. The main difference is that
the subsystems are no longer localized in disjoint regions of space-time. Instead, they are all
delocalized, and only when picking a pair of space-time points~$(x,y)$, the
phases in the sum~\eqref{PUU} determine to which subsystem the pair belongs.
Due to the obvious analogy to a hologram, we refer to this concept as
the {\em{holographic superposition of subsystems}} (but it does not seem to be
directly related to 't Hooft's holographic principle).
The main advantage of a holographic superposition is that a large number
of subsystems no longer leads to a ``fragmentation'' of space-time into disjoint space-time
regions. On the other hand, all the effects considered in this paper can be described
just as well by decoherent space-time regions. Therefore, the holographic superposition
is not essential for our purposes, and we shall not enter the detailed constructions here.

\appendix
\section{Settings where Entanglement is Impossible} \label{appA}
\subsection{Restriction to a Subsystem} \label{appA1}
In all realistic situations, the relevant measurements are performed only in a small subsystem of
the universe. Restricting attention to the subsystem, the effective fermionic state is no longer
of Hartree-Fock type, giving the hope that this might make it possible to describe entanglement.
The observation that the restriction to a subsystem gives more freedom to describe the
effective fermionic state was already used in~\cite[Appendix~A]{PFP} to show that
if only one-particle measurements are made, describing the system by a one-particle projector~$P$
is equivalent to the general Fock space formalism (this result corresponds to
Proposition~\ref{prp35} below). Here we generalize the methods to many-particle observables
and show that the restriction to a subsystem does {\em{not}} make it possible to describe general
entangled states.

In order to describe the subsystem, we assume that the one-particle observables
act only on a proper subspace~$\mathfrak{I} \subset \H$ (the ``inner system'').
Extending such an observable~$\cO$ by zero to all of~$H$,
we obtain a self-adjoint operator which vanishes on the orthogonal complement of~$\mathfrak{I}$.
Extending this operator by~\eqref{oneO} to the Fock space, 
we obtain an operator which preserves the number of particles and 
vanishes on the orthogonal complement of the space generated by~$\mathfrak{I}$,
\beq \label{mehrO}
\cO\::\: \F_n \rightarrow \F_n \text{ self-adjoint } \qquad \text{and} \qquad
\cO|_{\bra \:\Lambda_n(\mathfrak{I}^n) \:\ket^\perp} \equiv 0
\eeq
(for notational convenience the superscript of~$\cO^\F$ has been omitted).
In order to allow for many-particle observables, we can generalize~$\cO$ to be a general operator on~$\F$
which satisfies~\eqref{mehrO} and thus leaves the number of particles~$n$ fixed.

We again denote the projector on the corresponding $f$-particle Hartree-Fock state~$\Psi$
by~$P_f$ (see~\eqref{Ppure}, \eqref{wedge} and Proposition~\ref{prp13}),
\beq \label{Pfnew}
P_f = f! \: | \Psi \ra \la \Psi | \qquad \text{where} \qquad
\Psi = \psi_1 \wedge \cdots \wedge \psi_f \:.
\eeq
Setting~$O=\mathfrak{I}^\perp$ (the ``outer system''), we decompose all one-particle states~$\psi_i$ into their inner and outer parts,
\beq \label{inout}
\psi_i = \psi_i^\mathfrak{I} + \psi_i^O \qquad \text{with} \qquad
\psi_i^\mathfrak{I} \in \mathfrak{I},\: \psi_i^O \in O\:.
\eeq
Substituting this decomposition into~\eqref{Pfnew} and multiplying out, one gets a sum of
terms involving wedge products of the~$\psi_i^\mathfrak{I}$ and~$\psi_i^{O}$.
In order to keep track of the
combinatorics, it is convenient to denote by~$I$ a multi-index
\beq \label{multi1}
I=(i_1, \ldots, i_g) \qquad \text{with} \qquad 1 \leq i_1 < i_2 < \cdots < i_g \leq f \:,
\eeq
to set~$|I|:=g$, and to define the sign of~$I$ as the sign of a permutation
in~$\{1, \ldots, f\}$ which maps~$I$ to the set~$\{1, \ldots, g\}$, i.e.
\beq \label{multi2}
\sign( I ) = (-1)^{i_1 + \cdots + i_g + \frac{g(g+1)}{2}}\:.
\eeq
Furthermore, we denote the complement of~$I$ by~$O$, i.e.
\[ O=(j_1, \ldots, i_h) \qquad \text{with} \qquad 1 \leq j_1 < \cdots < j_h \leq f\:, \quad
g+h=f \:,\quad i_k \neq j_l \;\;\forall k,l\:. \]
Finally, we set
\[ \Psi^I = \psi^\mathfrak{I}_{i_1} \wedge \cdots \wedge \psi^\mathfrak{I}_{i_g}
\quad \text{and} \quad
\Psi^O = \psi^O_{j_1} \wedge \cdots \wedge \psi^O_{j_h} \:. \]
Using this notation, we obtain
\[ \Psi = \sum_{I} \sign(I)\: \Psi^I \wedge \Psi^O \:, \]
and thus we can write~\eqref{Pfnew} as follows,
\beq \label{PfIO}
P_f = \sum_{I, I'} f! \:\sign(I) \sign(I')\:
|\Psi^I \wedge \Psi^O \ra \la \Psi^{I'} \wedge \Psi^{O'} |\:.
\eeq

The inner system~$\mathfrak{I}$ is conveniently described by a density operator, which we
now define. A state~$\Psi \in \F$ (no matter if factorizable or entangled) is referred to as a
{\em{pure state}}.
Due to a limited knowledge on the physical system, the fermions can often not be described by a pure
state, but merely by an ensemble of states, coming with certain probabilities.
More precisely, one considers a family of pure states~$\Psi_1, \ldots, \Psi_L \in \F$ together with
corresponding probabilities~$p_1, \ldots, p_L$, normalized as follows,
\[ \|\Psi_l\|_{\F}=1 \qquad \text{and} \qquad 0 \leq p_l \leq 1\:,\quad \sum_{l=1}^L p_l = 1\:. \]
The expectation value of an observable~$\cO$ is then defined by
\beq \label{statcond}
\la \cO \ra = \sum_{l=1}^L p_l \:\la \Psi_l | \cO \Psi_l \ra \:.
\eeq
Introducing the operator
\[ \rho = \sum_{l=1}^L p_l \:|\Psi_l \ra \la \Psi_l | \:, \]
the expectation value can be expressed in analogy to~\eqref{expect} by
\beq \label{Trrho}
\la \cO \ra = \Tr (\rho \cO )\:,
\eeq
whereas the conditions~\eqref{statcond} become
\beq \label{rhocond}
\rho \geq 0 \qquad \text{and} \qquad \Tr(\rho)=1\:.
\eeq
The operator~$\rho$ is referred to as the {\em{density operator}}. If~$\rho$ is a projector,
it follows from~\eqref{rhocond} that~$\rho$ has rank one, and thus it can be written
in the form~\eqref{Ppure} with a pure state~$\Psi$. If~$\rho$ is not a projector, it is said
to describe a {\em{mixed state}}.

When computing expectation values of operators localized in our subsystem, we
can take the partial trace over~$O$ to obtain an equivalent description
of our quantum system by a density operator defined in the subsystem. This is made precise
in this next lemma.
\begin{Lemma} \label{lemma41} For any operator~$\cO$ on~$\F$ of the form~\eqref{mehrO},
the expectation value~\eqref{expect} can be expressed by~$\la \cO \ra = \Tr (
\rho\:\cO)$, where~$\rho$ is the density operator
\beq \label{stat}
\rho = \sum_{g=0}^f \; \sum_{\substack{I, I' \text{with} |I|=|I'|=g}}
g! (f-g)! \sign(I) \sign(I')\: \la \Psi^{O'} | \Psi^{O} \ra\:
|\Psi^I \ra \la \Psi^{I'}| \:.
\eeq
\end{Lemma}
\Proof Applying~\eqref{mehrO}, we find
\begin{align*}
\la \Psi^{I'} & \wedge \Psi^{O'} |\, \cO \left( \Psi^I \wedge \Psi^O \right) \ra
= \la \Psi^{I'} \wedge \Psi^{O'} | \left(\cO \Psi^I  \right) \wedge \Psi^O \ra \\
&=\la \Psi^{I'} \otimes \Psi^{O'} | \left(\cO \Psi^I  \right) \wedge \Psi^O \ra
= \frac{g! (f-g)!}{f!}\: \la \Psi^{O'} \:|\: \Psi^O \ra\:
\la \Psi^{I'} | \cO \Psi^I \ra \\
&= \frac{g! (f-g)!}{f!}\: \la \Psi^{O'} | \Psi^O \ra\:
\Tr \left(|\Psi^I \ra  \la \Psi^{I'} |\, \cO \right) .
\end{align*}
Using this relation in~\eqref{PfIO} gives the result.
\QED

We point out that the density operator~\eqref{stat} involves states of a variable number of
particles~$g=0,\ldots, f$. The coefficients depend on the inner
products~$\la \Psi^{O'} | \Psi^{O} \ra$ of the wave functions in the outer region. Since
in the outer region no measurements are possible, we cannot determine the wave
functions~$\psi_i^O$. Thus we can take the point of view that these wave functions
can be chosen arbitrarily. At first sight, this seems to give a lot of freedom to
choose the coefficients~$\la \Psi^{O'} | \Psi^{O} \ra$ in~\eqref{stat},
and thus one might conjecture that with~\eqref{stat} it should be possible to describe a general
entangled state. However, it is shown in Appendix~\ref{appA1} that this
conjecture is wrong, basically because many degrees of freedom in choosing
the wave functions~$\psi_i^O$ drop out when carrying out the sums in~\eqref{stat}.
We thus conclude that the restriction to a subsystem does {\em{not}} make it possible to
describe entangled states.

Let us analyze the density operator of Lemma~\ref{lemma41}.
In order to bring the inner products~$\la \Psi^O | \Psi^{O'} \ra$ 
into a more convenient form, we introduce the $(f \times f)$-matrix~$A$ by
\beq \label{Adef}
A_{ij} := \la \psi^O_i | \psi^O_j \ra \:;
\eeq
it can be considered as the Gram matrix of the one-particle wave functions in the outer system.
This matrix positive semi-definite matrix, but this property will not be used in what follows.
This has the advantage that our results apply even in a setting where the inner product~$\la .|. \ra$
is not positive definite (for example, one might think of a situation where the measurement scalar
product is indefinite on the microscopic scale).
The matrix~$A$ may be singular. Therefore, it is preferable to add a small multiple of the
identity matrix to obtain an invertible matrix. Thus for any~$\varepsilon>0$ we set
\beq \label{ABdef}
A^\varepsilon = A + \varepsilon\, \1 \qquad \text{and} \qquad B^\varepsilon = (A^\varepsilon)^{-1}\:.
\eeq
By~$B^\varepsilon_{I', I}$ we denote the $(g \times g)$-matrix obtained from~$B^\varepsilon$
by deleting all rows and columns except for those corresponding to~$I$ and~$I'$, respectively.

\begin{Thm} The expectation value~\eqref{expect} can be expressed by
\[ \la \cO \ra = \Tr ( \rho\,\cO) \:, \]
where~$\rho$ is the density operator
\beq \label{stat2}
\rho = \sum_{g=0}^f \;\sum_{\substack{I, I' \text{with} |I|=|I'|=g}}
g! \: \lim_{\varepsilon \searrow 0} \Big( \det(A^\varepsilon) \det(B^\varepsilon_{I',I}) \Big)
|\Psi^I \ra \la \Psi^{I'}| \:,
\eeq
and the matrices~$A^\varepsilon$, $B^\varepsilon$ are defined by~\eqref{Adef} and~\eqref{ABdef}.
\end{Thm}
\Proof Setting~$h=f-g$, the inner product in~\eqref{stat} is computed by
\beq \label{eqOpO}
\la \Psi^{O'} | \Psi^O \ra = \frac{1}{h!}
\sum_{\sigma \in S_h} \la \psi_{j'_{\sigma(1)}}^O
\otimes \cdots \otimes \psi_{j'_{\sigma(h)}}^O \:|\:
\psi_{j_1}^O \otimes \cdots \otimes \psi_{j_h}^O \ra
= \frac{1}{h!}\: \det(A_{O', O})\:, 
\eeq
where in the last step we used~\eqref{sdef} and the definition of the determinant.
Since the determinant is polynomial and thus continuous, it is obvious that
\[ \det(A_{O', O}) = \lim_{\varepsilon \searrow 0} \det(A^\varepsilon_{O', O})\:. \]
Using this relation in~\eqref{eqOpO} and~\eqref{stat}, we conclude that it remains to prove
for any invertible matrix~$A$ the identity
\beq \label{Cramer}
\det(A_{O', O}) = \sign(I') \sign(I) \:\det(A) \det(B_{I', I})\:,
\eeq
which relates the minors of~$A$ to the minors of its inverse.
For proving~\eqref{Cramer}, we first note that in the special case where~$I$ and~$I'$ consist of only one
index, this is the well-known Cramer's rule for the inverse of a matrix. In the general case, the
identity~\eqref{Cramer} is stated in~\cite[Section~0.8.4]{horn+johnson}. It can be proved
as follows. The signs in~\eqref{Cramer} can be understood from the fact that if rows or columns
of the matrices~$A$ and~$B$ are permuted without violating the ordering of the multi-indices~$I$,
$I'$, $O$ and~$O'$, then every such conjugation flips the sign of~$\det(A)$, and also the sign
of one of the functions~$\sign(I)$ or~$\sign(I')$. With such conjugations we can arrange that
\beq \label{multias}
I=I'=(1,\ldots, g) \qquad \text{and} \qquad O=O'=(g+1,\ldots, f)
\eeq
(but of course, the matrices~$A$ and~$B$ will no longer be Hermitian).
It remains to show that in this case,
\[ \det(A_{O', O}) = \det(A) \det(B_{I', I}) \:. \]

Using Laplace's formula, in case~\eqref{multias} the minor~$\det(A_{O',O})$ can 
be written as a multiple derivative of~$\det(A)$,
\beq \label{detOOp}
\det(A_{O',O}) = \frac{\partial}{\partial A_{11}} \cdots \frac{\partial}{\partial A_{gg}}
\det A \:.
\eeq
Using the standard formulas
\begin{align*}
\frac{\partial}{\partial A_{ii}} \det(A) &= \det(A) \,B_{ii} \\
\frac{\partial}{\partial A_{ii}} B_{jk} = \frac{\partial}{\partial A_{ii}} (A^{-1})_{jk}
&= -\left( B \left(\frac{\partial}{\partial A_{ii}} A\right) B \right)_{jk}
= -B_{ji} B_{ik}\:,
\end{align*}
one can iteratively carry out the derivatives in~\eqref{detOOp} to obtain
\[ \det(A_{O',O}) = \det(A) \times \Big(\text{homogeneous polynomial of degree~$g$
in $B_{jk}$
with~$j,k \in I$}\Big). \]
Going through the combinatorial details, one finds that this homogeneous polynomial coincides
precisely with~$\det(B_{I',I})$.
\QED

We now illustrate this theorem by a few examples and work out simple consequences.
We first explain how to recover a Hartree-Fock state.
\begin{Example} (Hartree-Fock state) {\em{ Choosing~$\Psi^\mathfrak{I}=\Psi$
and~$\Psi^O=0$, the Gram matrix~\eqref{Adef} vanishes and thus
\[ A^\varepsilon = \varepsilon \qquad \text{and} \qquad B^\varepsilon=\varepsilon^{-1}\:. \]
Hence
\[ \det(A^\varepsilon) \det(B^\varepsilon_{I',I}) = \varepsilon^{f-|I|}\: \delta_{I,I'}\:, \]
and this vanishes in the limit~$\varepsilon \searrow 0$ unless~$I=(1, \ldots, f)$. Hence~\eqref{stat2}
reduces to
\[ \rho = f! \, |\Psi \ra \la \Psi|\:, \]
so that the density operator coincides with the projector~$P_f$~\eqref{Pfnew} of the whole system. \\
\hspace*{1cm} \QEDrem }}
\end{Example}

\begin{Example} (A mixed state) {\em{ We choose~$f=3$, $\psi_1^{O}=0$,
$\|\psi_2^{O}\|=\|\psi_3^{O}\|=\kappa \in [0,1]$
and $\la \psi_2^{O} |\psi_3^{O} \ra = 0$. Thus
\[ A^\varepsilon = \diag \!\left(\varepsilon, \kappa^2+\varepsilon, \kappa^2+\varepsilon \right) \:, \]
and a short calculation yields that
\begin{align*}
\rho =& \kappa^4 | \psi^{\mathfrak{I}}_1 \ra \la \psi^{\mathfrak{I}}_1 | 
+ 2 \kappa^2 \left( | \psi^{\mathfrak{I}}_1 \wedge \psi^{\mathfrak{I}}_2 \ra
\la \psi^{\mathfrak{I}}_1 \wedge \psi^{\mathfrak{I}}_2 |  +
 | \psi^{\mathfrak{I}}_1 \wedge \psi^{\mathfrak{I}}_3 \ra
\la \psi^{\mathfrak{I}}_1 \wedge \psi^{\mathfrak{I}}_3 | \right) \\
&+ 3!\: | \psi^{\mathfrak{I}}_1 \wedge \psi^{\mathfrak{I}}_2 \wedge \psi^{\mathfrak{I}}_3 \ra
\la \psi^{\mathfrak{I}}_1 \wedge \psi^{\mathfrak{I}}_2 \wedge \psi^{\mathfrak{I}}_3 |\:.
\end{align*}
Thus the density operator of the subsystem is an ensemble of one, two and three particle states.
The one-particle state~$\psi^{\mathfrak{I}}_1$ is always occupied; this is because the
corresponding eigenvalue of the matrix~$A$ vanishes.
The two-particle component cannot be represented by a pure state; it is a mixed state.
}} \QEDrem
\end{Example}

We next show that when restricting attention to one-particle observables, the expectation value
of any vector in the fermionic Fock space can be approximated to arbitrary precision by
a one-particle projector~$P$.
\begin{Prp} \label{prp35} For every
normalized $n$-particle state~$\Psi$ in our subsystem, i.e.
\[ \Psi \in \overline{\bra \Lambda_n( {\mathfrak{I}}^n) \ket} \subset \F_n 
\qquad \text{and} \qquad \|\Psi\|=1 \:, \]
there is a sequence of projectors~$(P_k)_{k \in \N}$ in~$\H$ of finite rank~$f_k$, such that
for every one-particle observable~$\cO$,
\[ \la \Psi | \cO \Psi \ra = \lim_{k \rightarrow \infty} \Tr(P_k \cO)\:. \]
\end{Prp}
\Proof Writing~$\Psi$ as a linear combination of Hartree-Fock states and computing
the expectation value with~\eqref{oneO} and~\eqref{sdef}, we obtain
\[ \la \Psi | \cO \Psi \ra = \sum_{k,l} c_{k,l} \:\la \psi_k^{\mathfrak{I}} |
\cO \psi_l^{\mathfrak{I}} \ra \]
with suitable vectors~$\psi_k^{\mathfrak{I}} \in {\mathfrak{I}}$ and complex coefficients~$c_{k,l}$. 
Using an approximation argument, it clearly suffices to consider finite sums and finite linear combinations.
Choosing an orthonormal basis~$\psi_k$ of the subspace $\la \psi_k^{\mathfrak{I}} \ra \subset
{\mathfrak{I}}$, and expressing the~$\psi_k^{\mathfrak{I}}$ as linear combinations of
the~$\psi_k$, we obtain
\[ \la \Psi | \cO \Psi \ra = \sum_{k,l} \rho_{k,l} \:\la \psi_k |
\cO \psi_l \ra \:. \]
Diagonalizing the symmetric matrix~$\rho_{k,l}$ by a unitary transformation,
this representation simplifies to
\beq \label{exrep}
\la \Psi | \cO \Psi \ra = \sum_{k} \rho_k \:\la \psi_k |
\cO \psi_k \ra \:.
\eeq
Since the operator~$| \Psi \ra \la \Psi |$ is positive and normalized, it follows that
\[ \rho_k \geq 0 \qquad \text{and} \qquad \sum_k \rho_k = 1\:. \]

We choose an orthonormal family of vectors $(\phi_k)$ in $O$ and set
\[ \psi_k^\text{tot} = \sqrt{\rho_k}  \:\psi_k + \sqrt{1-\rho_k} \: \phi_k\:. \]
Then the family~$(\psi_k^\text{tot})$ is orthonormal, and thus
\[ P = \sum_k | \psi_k^\text{tot} \ra \la \psi_k^\text{tot} | \]
defines a projector in~$\H$. A short calculation shows that
the expectation value~$\Tr(P \cO)$ coincides with the right side of~\eqref{exrep}.
\QED

The next example shows that the restriction to a subsystem does {\em{not}} make it possible
to describe general entangled states.
\begin{Example} (The spatially separated singlet state) \label{ex36} {\em{
Let us try to realize the spatially separated singlet state of Example~\ref{exsinglet}.
We assume that the inner system~$\mathfrak{I} \subset \H$ has the orthonormal
basis~$(\psiu_A, \psid_B, \psid_A, \psiu_B)$.
Our goal is to find a projector~$P$ in~$\H$ such that the corresponding density
operator of the subsystem~\eqref{stat2} coincides with the projector onto the
singlet state~\eqref{singlet}, i.e.
\beq \begin{split}
\rho =& \: | \psiu_A \wedge \psid_B \ra \la \psiu_A \wedge \psid_B |
+ | \psid_A \wedge \psiu_B \ra \la \psid_A \wedge \psiu_B | \\
&- | \psiu_A \wedge \psid_B \ra \la \psid_A \wedge \psiu_B |
- | \psid_A \wedge \psiu_B \ra \la \psiu_A \wedge \psid_B | \:.
\end{split} \label{eproj}
\eeq
The simplest way to see that such a projector~$P$ does not exist is to observe that
in~\eqref{stat2} one necessarily gets the contribution involving four particles
\[  | \psiu_A \wedge \psid_B \wedge \psid_A \wedge \psiu_B \ra \la
\psiu_A \wedge \psid_B \wedge \psid_A \wedge \psiu_B |\:, \]
which is not present in~\eqref{eproj}. However, this argument is not fully convincing,
because an additional four-particle contribution would not be observable in the standard
spin correlation experiments where \textsc{Alice} and \textsc{Bob} can only detect one particle at a time.
Furthermore, by making the matrix elements of~$B^\varepsilon$ small, one could
try to arrange that the four-particle contribution is negligible. For these reasons, it
is preferable to show that we cannot even realize that the two-particle component of~\eqref{stat2}
coincides with~\eqref{eproj}, as we now explain.

Restricting attention to the two-particle component of~\eqref{stat2}, we need to consider
the determinants of the $2 \time 2$-submatrices of the matrix~$B^\varepsilon$.
Noting that in the case~$|I|=|I'|=2$, we can write the product of the determinants in~\eqref{stat2} as
\[ \det(A^\varepsilon) \det(B^\varepsilon_{I',I}) = 
\det \left( \sqrt{\det(A^\varepsilon)} \:B^\varepsilon_{I',I} \right) , \]
it suffices to consider the $2$-minors of the matrix
\[ B := \lim_{\varepsilon \searrow 0} \sqrt{\det(A^\varepsilon)} \:B^\varepsilon \:. \]
Grouping the first two and the last two basis vectors together, we write~$B$ as the block matrix
\[ B = \begin{pmatrix} B_{11} & B_{12} \\ B_{21} & B_{22} \end{pmatrix} , \]
whose entries are~$2 \times 2$ matrices acting on the subspaces~$\bra \{ \psiu_A, \psid_B \} \ket$
and~$\bra \psid_A, \psiu_B \ket$, respectively.
In order to realize~\eqref{eproj}, we need to arrange that the determinants of the submatrices~$B_{11}$
and~$B_{22}$ are non-zero, but all other $2$-minors must vanish.
Diagonalizing~$B_{11}$ and~$B_{22}$ by unitary transformations
in the subspaces~$\bra \{ \psiu_A, \psid_B \} \ket$ and~$\bra \{ \psid_A, \psiu_B \} \ket$, respectively,
we obtain
\[ B = \begin{pmatrix} \rho_1 & 0 & \overline{a} & \overline{c} \\
0 & \rho_2 & \overline{b} & \overline{d} \\
a & b & \rho_3 & 0 \\
c & d & 0 & \rho_4  \end{pmatrix} \qquad \text{with} \qquad \rho_i > 0 \:. \]
Evaluating the condition $\det B_{I,I'}=0$ for~$I=(1,3)$ and~$I'=(1,2)$, we see that~$b$ must
vanish. Similarly, taking $I=(1,4)$ and~$I'=(1,2)$ yields~$d=0$. Repeating this procedure
for~$I=(2,3)$ and~$I=(2,4)$ we find that~$a=c=0$.
Hence~$B$ is a positive diagonal matrix. But then the submatrix~$B_{I,I}$ with~$I=(1,3)$
has a non-zero determinant, a contradiction.
}} \QEDrem
\end{Example}

\begin{Remark} (Systems with purely classical bosonic fields) \label{remclassical} {\em{
Let us consider a system where second-quantized fermions are coupled to
classical bosonic fields, in the pure sense that also all measurement devices
can be described by the classical field equations.
As pointed out in~\cite[Appendix~A]{PFP}, the coupling of the fermions to
classical fields can be described by one-particle observables
(like the expectation value of the Dirac current in Maxwell's equations
or the expectation value of the energy-momentum tensor in Einstein's equations).
In view of Proposition~\ref{prp35}, the resulting
system can be described by a one-particle projector~$P$.
On the other hand, we saw in Example~\ref{ex36} that this framework does not allow for
the description of general entangled states. We conclude that
the physical phenomenon of entanglement makes it necessary to consider
second quantized bosonic fields.

This consideration suggests that there is a close connection between entanglement and
second quantization of the bosonic fields. This connection becomes clearer in
Chapters~\ref{secmix} and~\ref{secbosonic},
because the method of microscopic mixing, which allows for
the description of entanglement (see Chapter~\ref{secmix}), also
makes it possible to describe second quantized bosonic fields (see Chapter~\ref{secbosonic}).
}} \QEDrem
\end{Remark}

\subsection{Microscopic Mixing of the Wave Functions} \label{appA2}
In order to analyze the setting of Section~\ref{secmm} in more detail, it is useful to get
a connection to the fermionic Fock space formalism.
For a one-particle wave function~$\phi \in \H$, the {\em{creation operator}}~$a^\dagger(\phi)$ is
defined on Hartree-Fock states by
\[ a^\dagger(\phi) \left( \psi_1 \wedge \cdots \wedge \psi_n \right) =
\phi \wedge \psi_1 \wedge \cdots \wedge \psi_n\:. \]
By linearity, it is extended to an operator $a^\dagger(\phi) : \F_n \rightarrow \F_{n+1}$.
Its adjoint $a(\phi) : \F_{n+1} \rightarrow \F_n$, the so-called {\em{annihilation operator}},
acts on Hartree-Fock states by
\[ a(\phi) \left( \psi_1 \wedge \cdots \wedge \psi_{n+1} \right) =
\sum_{k=1}^{n+1} (-1)^{k+1} \:\la \phi | \psi_k \ra
\:\psi_1 \wedge \cdots \wedge \psi_{k-1} \wedge \psi_{k+1}
\wedge \cdots \wedge \psi_{n+1}\:. \]
A straightforward calculation shows that the creation and annihilation operators satisfy the
anti-commutation relations
\beq \label{anticommute}
\left\{ a(\phi), a^\dagger(\psi) \right\} = \la \phi | \psi \ra\:\1_{\F} \:.
\eeq
A one-particle observable~\eqref{oneO} can be expressed in terms of the creation and
annihilation operators by
\[ \cO^\F = \sum_{k,l} a^\dagger(\phi_k) \:\la \phi_k | \cO \phi_l \ra \:a(\phi_l)\:, \]
where~$(\phi_l)$ denotes an orthonormal basis of~$(\H, \la .|. \ra)$.
Products of one-particle observables can be transformed with the anti-commutation
rule~\eqref{anticommute}; for example,
\begin{align}
\cO_1^\F \cO_2^\F =& \sum_{k_1,k_2,l_1, l_2}
a^\dagger(\phi_{k_1}) \,a^\dagger(\phi_{k_2}) \:
\la \phi_{k_1} | \cO_1 \phi_{l_1} \ra \la \phi_{k_2} | \cO_2 \phi_{l_2} \ra\:
a(\phi_{l_1}) \,a(\phi_{l_2}) \\
&+\sum_{k,l} a^\dagger(\phi_k) \:\la \phi_k | \cO_1 \cO_2 \phi_l \ra \:a(\phi_l)\:, \label{wi2}
\end{align}
where in the last line we used the completeness of the basis~$(\phi_l)$.

A useful rule in quantum field theory is {\em{Wick ordering}}, denoted by colons, which states
that all creation operators should be written to the left and all annihilation operators to the right,
leaving out all terms which would be generated by the anti-commutations.
For example, Wick ordering the above product of one-particle observables amounts to
omitting the term~\eqref{wi2},
\beq \label{wi3}
:\! \cO_1^\F \cO_2^\F \!: \;= \sum_{k_1,k_2,l_1, l_2}
a^\dagger(\phi_{k_1}) \,a^\dagger(\phi_{k_2}) \:
\la \phi_{k_1} | \cO_1 \phi_{l_1} \ra \la \phi_{k_2} | \cO_2 \phi_{l_2} \ra\:
a(\phi_{l_1}) \,a(\phi_{l_2}) \:.
\eeq
A general Wick-ordered two-particle observable can be written as
\beq \label{twopart}
:\!{\mathcal{O}}\!: \;= \sum_{k_1,k_2,l_1, l_2}
a^\dagger(\phi_{k_1}) \,a^\dagger(\phi_{k_2}) \:
g(k_1,k_2,l_1,l_2)\:
a(\phi_{l_1}) \,a(\phi_{l_2}) \:,
\eeq
where the function~$g$ is anti-symmetric in its first and last two arguments,
\beq \label{twosymm}
g(k_1,k_2,l_1,l_2) = -g(k_2,k_1,l_1,l_2) = -g(k_1,k_2,l_2,l_1)\:.
\eeq
In the next proposition we express the expectation value of this two-particle observable
in terms of the one-particle projector~$P$.
\begin{Prp} \label{prp25}
Describing a many-fermion state by a projector~$P$ in~$(\H, \la .|. \ra)$,
the expectation value of the two-particle operator~\eqref{twopart}
with~$g$ according to~\eqref{twosymm} is given by
\[ \la :\!{\mathcal{O}}\!: \ra
= \sum_{k \neq l} \Big( \la \phi_k | P \phi_k \ra \la \phi_l | P \phi_l \ra
- \la \phi_k | P \phi_l \ra \la \phi_l | P \phi_k \ra \Big) \,g(k,l,k,l)\:. \]
The expectation value of the Wick ordered product~\eqref{wi3} is
\beq \label{wiex}
\la :\!\cO_1^\F \cO_2^\F \!: \ra =
\Tr_\H (P \cO_1 ) \Tr_\H (P \cO_2) - \Tr_\H (P \cO_1 P \cO_2)\:.
\eeq
\end{Prp}
\Proof We again represent the projector~$P_f$ in the form~\eqref{Ppure}.
Using the obvious transformation law of the function~$g$ under basis transformations,
we can arrange that~$\phi_1=\psi_1$, \ldots, $\phi_f=\psi_f$. Then the results follows
from a straightforward calculation.
\QED
Note that the formula~\eqref{wiex} coincides with~\eqref{tr2}, except that the
summand~$\Tr_\H (P \cO_1 \cO_2)$ is now missing as a consequence of the Wick ordering.

\begin{Prp} \label{prpnosinglet}
In the setting of Section~\ref{secmm}, there is no projector~$P$ in $\H$ which reproduces the expectation
values of the spatially separated singlet state~\eqref{singlet} with respect to the
spin operators of  \textsc{Alice} and \textsc{Bob} and the corresponding two-particle spin correlation operators.
\end{Prp}
\Proof We let~$S_\uparrow$ be the spin operator having the
expectation values one if the spin is up and zero if the spin is down.
Similarly, $S_\downarrow$ is the operator for spin down.
The spin operators corresponding to the observers \textsc{Alice} and \textsc{Bob} are
denoted by~$S_{\uparrow \!/\! \downarrow,A}$ and~$S_{\uparrow \!/\! \downarrow,B}$, respectively;
they are operators in~$\H$.
Taking for convenience the Wick ordered products of the corresponding operators on the
Fock as defined by~\eqref{wi3}, we can compute the following expectation values of the
singlet state~\eqref{singlet},
\beq \label{onetwo}
\begin{split}
&\la S_{\uparrow,A} \ra = \la S_{\downarrow,B} \ra = \frac{1}{2} \\
\la :\! S_{\uparrow,A}^\F \, S_{\downarrow,B}^\F \!: \ra = \frac{1}{2}&\:,\quad
\la :\! S_{\uparrow,A}^\F \,S_{\uparrow,A}^\F \!: \ra = \la :\! S_{\downarrow,B}^\F \,
S_{\downarrow,B}^\F \!: \ra = 0
\end{split}
\eeq
(all the other expectation values are irrelevant for the proof).
Assuming that the statement of the proposition
is false, these expectation values can be reproduced by a suitable
one-particle projector~$P$. Applying Lemma~\ref{lemma24} and Proposition~\ref{prp25}, we can
then express the expectation values by traces over the one-particle Hilbert space~$\H$,
\[ \la {\mathcal{O}}^\F \ra = \Tr({P \mathcal{O}}) \quad \text{and} \quad
\la :\!{\mathcal{O}}_1^\F {\mathcal{O}}_2^\F \!: \ra =
\Tr(P {\mathcal{O}}_1) \Tr(P {\mathcal{O}}_2) - \Tr(P {\mathcal{O}}_1 P {\mathcal{O}}_2)\:. \]
Using the idempotence of~$P$, the arguments of these traces can all be rewritten purely in
terms of the operators
\beq \label{Tdef}
T_{\uparrow,A} := P S_{\uparrow,A} P \qquad \text{and} \qquad
T_{\downarrow,B} := P S_{\downarrow,B} P \:,
\eeq
which are symmetric and of finite rank.
The relations~\eqref{onetwo} give rise to the conditions
\beq \label{singcond}
\Tr(T_{\uparrow,A}) = \Tr(T_{\downarrow,B}) = \frac{1}{2}\:,\quad
\Tr(T_{\uparrow,A}^2) = \Tr(T_{\downarrow,B}^2) = \frac{1}{4}\:,\quad
\Tr(T_{\uparrow,A} T_{\downarrow, B}) = -\frac{1}{4}\:.
\eeq
At this point it is helpful to regard the operators~$T_{.,.}$ as vectors in the real
Hilbert space of symmetric Hilbert-Schmidt operators with the scalar
product~$\la A,B \ra_{HS} = \Tr(A B)$ and corresponding norm~$\|A\|_{HS}
= \sqrt{\la A,A \ra_{HS}}$. Then the last two equations in~\eqref{singcond} imply that
\beq \label{HS}
\| T_{\uparrow,A} + T_{\downarrow,B} \|_{HS}^2 =
\Tr(T_{\uparrow,A}^2) + 2 \: \Tr(T_{\uparrow,A} T_{\downarrow, B}) + \Tr(T_{\downarrow,B}) = 0\:.
\eeq
It follows that~$T_{\uparrow,A}= -T_{\downarrow,B}$, in contradiction to the first
equation in~\eqref{singcond}.
\QED
We point out that in this proof we did not use that the one-particle
operators are invariant on the subsystems~\eqref{ainv}. Thus dropping this assumption
would not change the statement of Proposition~\ref{prpnosinglet}.

\Thanks{{\em{Acknowledgments:}} I would like to thank Andreas Grotz, Mario Kieburg, Joel Smoller,
Roderich Tumulka, Tilo Wettig and the referees for valuable comments on the manuscript.}

%\bibliographystyle{amsplain}
%\bibliography{../felix}

\def\dbar{\leavevmode\hbox to 0pt{\hskip.2ex \accent"16\hss}d}
\providecommand{\bysame}{\leavevmode\hbox to3em{\hrulefill}\thinspace}
\providecommand{\MR}{\relax\ifhmode\unskip\space\fi MR }
% \MRhref is called by the amsart/book/proc definition of \MR.
\providecommand{\MRhref}[2]{%
  \href{http://www.ams.org/mathscinet-getitem?mr=#1}{#2}
}
\providecommand{\href}[2]{#2}

\end{document}